\documentstyle[12pt,pstricks,pst-node,pst-coil]{article}
\newcommand{\quattrova}{($\phi^{\scriptscriptstyle a},
c^{\scriptscriptstyle a},
\lambda_{\scriptscriptstyle  a},{\bar c}_{\scriptscriptstyle a}$)~}

\newcommand{\be}{\begin{equation}}
\newcommand{\ee}{\end{equation}}
\newcommand{\bea}{\begin{eqnarray}}
\newcommand{\eea}{\end{eqnarray}}

\def \HT{{\widetilde{\mathcal H}}}
\def \LT{{\widetilde{\mathcal L}}}
\def \KT{{\widetilde{\mathcal K}}}
\def \DT{{\widetilde{\mathcal D}}}

\def \HCT{{\widehat{\widetilde{\mathcal H}}}}
\def \KCT{{\widehat{\widetilde{\mathcal K}}}}
\def \DCT{{\widehat{\widetilde{\mathcal D}}}}

\def \QH{{Q_{\scriptscriptstyle H}}}
\def \QK{{Q_{\scriptscriptstyle K}}}
\def \QD{{Q_{\scriptscriptstyle D}}}

\def \QBH{{\overline{Q}}_{\scriptscriptstyle H}}
\def \QBK{{\overline{Q}}_{\scriptscriptstyle K}}
\def \QBD{{\overline{Q}}_{\scriptscriptstyle D}}
\def \Qb{ Q_{\scriptscriptstyle BRS}}
\def \QBb{{\overline {Q}}_{\scriptscriptstyle BRS}}

\def \QCH{{\widehat{Q}_{\scriptscriptstyle H}}}
\def \QCK{{\widehat{Q}_{\scriptscriptstyle K}}}
\def \QCD{{\widehat{Q}_{\scriptscriptstyle D}}}
\def \QCBH{{\widehat{\overline{Q}}}_{\scriptscriptstyle H}}
\def \QCBK{{\widehat{\overline{Q}}}_{\scriptscriptstyle K}}
\def \QCBD{{\widehat{\overline{Q}}}_{\scriptscriptstyle D}}
\def \QCb{{\widehat Q}_{\scriptscriptstyle BRS}}
\def \QCBb{{\widehat{\overline {Q}}}_{\scriptscriptstyle BRS}}
\def \QDT{Q^t_{\scriptscriptstyle D}}
\def \QBDT{\overline{Q}^t_{\scriptscriptstyle D}}
\def \QKT{Q^t_{\scriptscriptstyle K}}
\def \QBKT{\overline{Q}^t_{\scriptscriptstyle K}}
\def \NDT{ N^{t}_{\scriptscriptstyle D}}
\def \NKT{ N^{t}_{\scriptscriptstyle K}}

\def \QS{{Q_{\scriptscriptstyle S}}}

\def \NH{ N_{\scriptscriptstyle H}}
\def \NK{ N_{\scriptscriptstyle K}}
\def \ND{ N_{\scriptscriptstyle D}}

\def \HS{H_{\scriptscriptstyle SUSY}}

\newcommand{\scite}{~\cite}

\begin{document}

\baselineskip =15.5pt
\pagestyle{plain}
\setcounter{page}{1}

\begin{titlepage}

\begin{flushright}
\end{flushright}
\vfil

\begin{center}
{\huge A New Superconformal Mechanics.}
\end{center}

\vfil

\begin{center}
{\large E.Deotto, G.Furlan, E.Gozzi}\\
\vspace {1mm}
Dipartimento di Fisica Teorica, Universit\`a di Trieste, \\
Strada Costiera 11, P.O.Box 586, Trieste, Italy \\ and INFN, Sezione 
di Trieste.\\
\vspace {1mm}
\vspace{3mm}
\end{center}

\vfil

\begin{center}
{\large Abstract}
\end{center}

\noindent
In this paper we propose a new supersymmetric extension of conformal mechanics. 
The Grassmannian variables that we introduce  are the basis of the forms and
of the vector-fields built over the symplectic
space of the original system. Our supersymmetric
Hamiltonian itself turns out to have a clear geometrical meaning being the  
Lie-derivative of the Hamiltonian flow of  conformal mechanics.  
Using superfields we derive a constraint which gives the exact solution 
of the supersymmetric system in a way analogous
to the constraint in configuration space which solved the original 
non-supersymmetric model. Besides the supersymmetric extension of the
original Hamiltonian, we also provide the extension of the other
conformal generators present in the original system. These extensions
have also a supersymmetric character being the square of some Grassmannian
charge. We build the whole superalgebra of these charges and analyze their
closure. The representation of the even part of this superalgebra
on the odd part turns out to be integer and not spinorial in character.
\vfil
\end{titlepage}
\newpage
\section{Introduction}
More than 20 years ago a conformally-invariant quantum mechanical model
was proposed and studied in ref.[1]. Recently that model has attracted 
again some interest  in connection with black-holes. It has been proven\scite{KAL} 
in fact that the dynamics of a particle near the horizon 
of an extreme Reissner-Nordstr\o m black-hole is governed in its radial motion
by the Lagrangian of ref.\scite{DFF}. This can be considered a manifestation
at the quantum mechanical level of the correspondence between Gravity
on $AdS$ and conformal field theory\scite{MAL}. In this case it is the
correspondence $AdS_{\scriptscriptstyle 2}/CFT_{\scriptscriptstyle 1}$.

In 1983/1984 two groups\scite{FUB} independently made a supersymmetric
generalization of conformal mechanics. It has been proved recently\scite{KAL} that
the motion of a {\it super}-particle near an extreme Reissner-Nordstr\o m 
black-hole is governed by a ``relativistic" generalization  of the
supersymmetric conformal mechanics proposed in ref.\scite{FUB}.
So also  the supersymmetric version of conformal mechanics seems to
hold some interest for black-hole physics. For these conformal models
(both for the original\scite{DFF}~and for the
supersymmetric one\scite{FUB})~a nice geometrical approach was pioneered 
in ref.\scite{RUS}. 

In this paper we will build a new supersymmetric extension of conformal
mechanics. It is tailored on a path-integral approach to classical mechanics
developed in ref.\scite{ENNIO}. The Grassmannian variables which
appear in this formalism are the basis of the forms\scite{MARS} and 
vector fields which one can build over the symplectic space of the 
original conformal system. The weight in this path-integral provides 
what is known as the Lie-derivative\scite{MARS} of the Hamiltonian flow. 
This Lie-derivative turns out to be supersymmetric\scite{ENNIO}.

The reader may ask which is the difference between our extension and the
one of ref.\scite{FUB} which was tailored on the supersymmetric quantum
mechanics of Witten\scite{WITTE}.
Basically the authors of ref.\scite{FUB} took the original conformal
Hamiltonian and added a Grassmannian part in order to make the whole Hamiltonian
supersymmetric. Our procedure and extension is different and more geometrical 
as will be explained later on in the paper. One difference for example
is that while the equations of motion for our bosonic variables are
the same as those of the old conformal model\scite{DFF}, the analog
equations  of  ref.\scite{FUB} have an extra piece. Another difference is that, 
once we use the language of superfields, while we have to stick  
the superfield into the old conformal potential\scite{DFF} the authors of 
ref.\scite{FUB} have to stick
it into a potential which is {\it derived} from the old conformal potential.

The paper is organized in the following manner: In section {\bf 2} we give a 
very brief outline of conformal mechanics\scite{DFF} and of the
supersymmetric extension present in the literature\scite{FUB}; 
in section {\bf 3} we put forward our 
supersymmetric extension and explain its geometrical
structure. In the same section  we build a whole set of charges connected 
with our extension and study their algebra in details. 
In section {\bf 4} we show that, differently
from the superconformal algebra of\scite{FUB} where the even part has
a spinorial representation on the odd part, ours is a non-simple
superalgebra whose even part has a reducible and  integer representation 
on the odd part. Nevertheless we can call it a superconformal
algebra because we have the charges which result from a combined  supersymmetry
and  conformal transformation. We think that the usual idea\scite{KAC}
which says that a superconformal algebra has its even part represented
spinorially on the odd one applies only to relativistic systems and not
to ours.
In section {\bf 5} we provide a superspace version for our model
and for the whole set of charges. In doing that we find a new {\it universal}
charge which is present always in our formalism
\cite{ENNIO} even for non-conformal model. This new charge 
acts on the variables in such a manner as to rescale the  over-all Lagrangian.
This charge is  not present in supersymmetric quantum mechanics\scite{FUB}.
In section {\bf 6}, using superfield variables, we provide an exact solution
of our model by giving a constraint in superfield space analogous to the
one found for configuration space in\scite{DFF} and which provided the
solution of the original conformal system. We confine some calculations
to a couple of  appendices.

In this paper we do not have applications of our 
supersymmetric extension of conformal mechanics to black-hole physics.
We thought anyhow worth to write up these mathematical results
because we feel that something deeply geometrical is behind
the connection conformal mechanics/black-holes or in general
the correspondence {\it AdS/CFT}. To grasp these geometrical issues it
is better to use  models where geometry is manifest and beautiful
and we believe ours has these features. We leave to others the task of
finding possible connection of our supersymmetric extension with black-holes.
\section{Conformal Mechanics  and its Supersymmetric \break Extension}

In this section we will briefly review conformal mechanics\scite{DFF}
and its standard supersymmetric extension\scite{FUB}.

The Lagrangian proposed in\scite{DFF} is

\be
\label{eq:uno}
L=\frac{1}{2}\left[{\dot q}^{2}-{g\over q^{2}}\right].
\ee

\noindent It is easy to prove that this Lagrangian is invariant under the
following transformations:

\newpage

\bea
\label{eq:due}
t^{\prime} & = & {{\alpha t+\beta}\over {\gamma t + \delta}};\\
\label{eq:tre}
q^{\prime}(t^{\prime}) & = & {q(t)\over (\gamma t +\delta)};\\
\label{eq:quattro}
\nonumber\\
\mbox{with} && \alpha\delta-\beta\gamma=1;
\eea

\noindent which are nothing else than the conformal transformations in 0+1 dimensions.
They are made of the combinations of the following three transformations:

\bea
\label{eq:cinque}
t^{\prime} & = & \alpha^{2}t~~~~~~~\mbox{\it dilations},\\
\label{eq:sei}
t^{\prime} & = & t+\beta ~~~~~\mbox{\it time-translations},\\
\label{eq:sette}
t^{\prime} & = & {t\over {\gamma t+1}}~~~ \mbox{\it special-conformal~transformations.}
\eea

\noindent The Noether charges associated\scite{DFF} to these three symmetries are:

\bea
\label{eq:otto}
H & = & \frac{1}{2}\left(p^{2}+{g\over q^{2}}\right);\\
\label{eq:nove}
D & = & tH-{1\over 4}(qp+pq);\\
\label{eq:dieci}
K & = & t^{2}H-\frac{1}{2}t(qp+pq)+\frac{1}{2}q^{2}.
\eea

\noindent Using the quantum commutator~$[q,p]=i$, the algebra of the three Noether
charges above is:

\bea
\label{eq:undici}
[H,D] & = & iH;\\
\label{eq:dodici}
[K,D] & = & -iK;\\
\label{eq:tredici}
[H,K] & = & 2iD.
\eea

\noindent The fact that $D$ and $K$ do not commute with $H$ does not mean that
they are not conserved. In fact what is true is that, as the $H,D,K$
above are explicitly dependent on $t$, we have:

\bea
\label{eq:quattordici}
&\displaystyle{\partial D\over \partial t}\neq 0;~~~~~~{\partial K\over\partial t}\neq 0;&\\
\label{eq:quindici}
&\displaystyle\frac{dD}{dt}=\frac{dK}{dt}=0.&
\eea

As the $H,D,K$ are conserved, their expressions at $t=0$ which are\footnote{The RHS of 
eqs.(16)-(18) is understood with $p$ and $q$ at $t=0$ even if we do not put any subindex
$(.)_0$ on them. Moreover $H_0$ and $H$ have even the same functional form and so we will drop 
the subindex $(.)_0$ on $H$.}:

\bea
\label{eq:sedici}
H_{0} & = & \frac{1}{2}\left[p^{2}+{g\over q^{2}}\right],\\
\label{eq:diciassette}
D_{0} & = & -{1\over 4}\left[qp+pq\right],\\
\label{eq:diciotto}
K_{0} & = & \frac{1}{2}q^{2},
\eea

\noindent satisfy the same algebra as those at time $t$ (see eqs. (\ref{eq:undici})-(\ref{eq:tredici})). 
This algebra is $SO(2,1)$ which is 
known\scite{WYB} to be isomorphic to the conformal group in $0+1$ dimensions.

Let us now turn to the supersymmetric extension of this model
proposed in ref.\scite{FUB}.
The Hamiltonian is:

\be
\label{eq:diciannove}
\HS = \frac{1}{2}\left(p^{2}+{g\over q^{2}}+{\sqrt {g}\over
q^{2}}[\psi^{\dag},\psi]_{\scriptscriptstyle -}\right)
\ee
\noindent
where $\psi,\psi^{\dag}$ are Grassmannian variables whose anticommutator
is $[\psi,\psi^{\dag}]_{\scriptscriptstyle +}=1$. As one can notice, in $\HS$
there is a first bosonic piece which is the conformal Hamiltonian of 
eq.(\ref{eq:uno}), plus a Grassmannian part. Note that the equations
of motion for ``$q$" have an extra piece with respect to the equations
of motion of the old conformal mechanics\scite{DFF}.

To make contact with supersymmetric quantum mechanics\scite{WITTE}
let us notice that $\HS$ can be written as:

\bea
\label{eq:venti}
\HS & = & \frac{1}{2}\left[Q,Q^{\dag}\right]_{\scriptscriptstyle +}=\\
\label{eq:ventuno}
 & = & \frac{1}{2}\left(p^{2}+\left({dW\over dq}\right)^{2}-[\psi^{\dag},\psi
]_{\scriptscriptstyle -}{d^{2}W\over dq^{2}}\right)
\eea

\noindent where the supersymmetry charges are:

\bea
\label{eq:ventidue}
&&Q=\psi^{\dag}\left(-ip+{dW\over dq}\right),\\
\label{eq:ventitre}
&&Q^{\dag}=\psi\left(ip+{dW\over dq}\right),
\eea

\noindent and $W$ is the superpotential which, in this case of conformal mechanics,
turns out to be:

\be
\label{eq:ventiquattro}
W(q)=\sqrt{g}\log q.
\ee

\noindent It is interesting to see what we obtain when we combine a supersymmetric
transformation with a conformal one generated by the $(H,K,D)$
elements of the $SO(2,1)$ algebra (\ref{eq:undici}),(\ref{eq:dodici}),(\ref{eq:tredici}).
We get what is called a {\it superconformal} transformation. In order to understand 
this better let us list the following eight operators:
\vskip 1cm
\begin{center}
{\bf TABLE 1} 
\end{center}

\[
\begin{array}{|rcl|}
\hline && \\
H & = &\displaystyle\frac{1}{2}\Bigl[p^{2}+{{g+2\sqrt{g}B}\over q^{2}}\Bigr];\\
D & = &\displaystyle -{[q,p]_{\scriptscriptstyle +}\over 4};\\
K & = &\displaystyle{q^{2}\over 2};\\
B & = &\displaystyle{[\psi^{\dag},\psi]_{\scriptscriptstyle -}\over 2};\\
Q & = &\displaystyle \psi^{\dag}\Bigl[-ip+{\sqrt{g}\over q}\Bigr];\\
Q^{\dag} & = &\displaystyle \psi\Bigl[ip+{\sqrt{g}\over q}\Bigr];\\
S & = &\displaystyle \psi^{\dag}q;\\
S^{\dag} & = &\displaystyle \psi q.\\
&& \\
\hline
\end{array}
\]

The algebra of these operators is closed and given in the table below:

\begin{center}
{\bf TABLE 2}
\end{center}

\[
\begin{array}{|lll|}
\hline & & \\

[H,D]=iH; \hspace{3cm}&[K,D]=-iK; \hspace{3cm} &[H,K]=2iD; \\

[Q,H]=0; &[Q^{\dag},H]=0; &[Q,D]={i\over 2}Q; \\

[Q^{\dag},K]=S^{\dag}; &[Q,K]=-S; &[Q^{\dag},D]={i\over 2}Q^{\dag}; \\

[S,K]=0; &[S^{\dag},K]=0; &[S,D]=-{i\over 2}S; \\

[S^{\dag},D]=-{i\over 2}S^{\dag}; &[S,H]=-Q; &[S^{\dag},H]=Q^{\dag}; \\

[Q,Q^{\dag}]=2H; &[S,S^{\dag}]=2K; & \\

[B,S]=S; &[B,S^{\dag}]=-S^{\dag}; & \\
 
[Q,S^{\dag}]=\sqrt{g}-B+2iD; &[B,Q]=Q; &[B,Q^{\dag}]=-Q^{\dag}; \\

& & \\
\hline
\end{array} 
\]

\vskip 1cm
\noindent all other commutators are zero or derivable from these by Hermitian conjugation.
The square-brackets $[(.),(.)]$ in the algebra above 
are  {\it graded}-commutators and from now on we shall
not put on them the subindex $+$ or $-$ as we did before. They are commutators
or anticommutators according to the Grassmannian nature of the operators 
entering the brackets.

As it is well known a {\it superconformal} transformation is a combination
of a supersymmetry transformation and a conformal one. We see from the
algebra above that the commutators of the supersymmetry 
generators $ (Q,Q^{\dag}) $ with the three conformal generators 
$ (H,K,D) $ generate
a new operator which is $ S $. Including this new one we generate an algebra
which is closed  provided that we introduce the operator $B$ of {\bf TABLE 3}. This is the last 
operator we need. 

\section{A New Supersymmetric Extension of Conformal Mechanics}

In this section we are going to present a new supersymmetric extension
of conformal mechanics. This extension is tailored on a path-integral
approach to classical mechanics (CM) developed in ref.\scite{ENNIO}.
The idea is to give a {\it path integral} for CM whose operatorial counterpart
be the well-known   {\it operatorial} version of CM  as given
by the {\it Liouville}  operator\scite{KOOP}.
We will be brief here because more details can be found in\scite{ENNIO}.

Let us start with a $2n$-dimensional phase space ${\cal M}$ whose coordinates
are indicated as $\phi^{a}$ with $a=1,\ldots,2n$, i.e.: $\phi^{a}=(q^1,\ldots,
q^n;p^1,\ldots,p^n)$. Let us write the 
Hamiltonian of the system as $H(\phi)$ and the symplectic-matrix as
$\omega^{ab}$.
The equations of motion are then:

\be
\label{eq:venticinque}
{\dot\phi }^{a}=\omega^{ab}{\partial H\over\partial\phi ^{b}}.
\ee

\noindent We shall put forward,  as path integral for CM, one that forces all paths
in ${\cal M}$
to sit on the classical ones. The {\it classical} analog of the quantum generating 
functional is:

\be
\label{eq:ventisei}
Z_{\scriptscriptstyle CM}[J]=N\int{\cal D}\phi~\tilde{\delta}[\phi
(t)-\phi _{cl}(t)]\exp\left[\int J\phi~dt\right] 
\ee 

\noindent where $\phi$ are the $\phi^{a}\in{\cal M}$, $\phi_{cl}$ are the
solutions of eq.(\ref{eq:venticinque}),
$J$ is an external current and $\widetilde{\delta}[.]$ is a functional
Dirac delta which forces
every path $\phi(t)$ to sit on a classical one $\phi_{cl}(t)$. Of course
there are all
possible initial conditions integrated over in eq.(\ref{eq:ventisei}).

We should now check if the path integral of eq.(\ref{eq:ventisei}) leads 
to the operatorial 
formulation\scite{KOOP} of CM. To do that let us first
rewrite the functional Dirac delta in eq.(\ref{eq:ventisei}) as:

\begin{equation}
\label{eq:ventisette}
{\tilde\delta}[\phi -\phi _{cl}]={\tilde\delta}[{\dot\phi
^{a}-\omega^{ab}
\partial_{b}H]~det [\delta^{a}_{b}\partial_{t}-\omega^{ac}\partial_{c}\partial
_{b}H}]
\end{equation}

\noindent where we have used the functional analog of the relation
~$\delta[f(x)]=\frac{\delta[x-x_i]}{\Bigm|\frac{\partial f}{\partial
x}\Bigm|_{x_i}}$.
The determinant which appears in eq.(\ref{eq:ventisette}) is always 
positive and so we can drop 
the modulus sign $|.|$. 
The next step is to insert eq.(\ref{eq:ventisette}) in eq.(\ref{eq:ventisei}) 
and write the $\tilde{\delta}[.]$
as a Fourier transform over some new variables $\lambda_{a}$, i.e.:

\begin{equation}
\label{eq:ventotto}
{\tilde{\delta}}\left[{\dot\phi }^{a}-\omega^{ab}{\partial
H\over\partial\phi ^{b}}\right]=
\int{\cal D}\lambda_{a}\exp\left\{i\int\lambda_{a}\left[{\dot \phi
}^{a}-\omega^{ab}
{\partial H\over\partial\phi ^{b}}\right]dt\right\}.
\end{equation}

\noindent We then express the determinant
$det[\delta^{a}_b\partial_t-\omega^{ac}\partial_c\partial_bH]$ via
Grassmannian variables $\bar{c}_a, c^{a}$ as:

\begin{equation}
\label{eq:ventinove}
det[\delta^{a}_{b}\partial_{t}-\omega^{ac}\partial_{c}\partial_{b}H]
=\int{\cal D}c^{a}{\cal D}{\bar c}_{a}\exp\left\{-\int {\bar c}_{a}[\delta^{a}_{b}
\partial_{t}-\omega^{ac}\partial_{c}\partial_{b}H]c^{b}dt\right\}.
\end{equation} 

\noindent Inserting the RHS of eqs. (\ref{eq:ventotto})(\ref{eq:ventinove}) in eq.(\ref{eq:ventisette}) 
and then in eq.(\ref{eq:ventisei}) we get:

\begin{equation}
\label{eq:trenta}
Z_{\scriptscriptstyle CM}[0]=\int{\cal D}\phi ^{a}{\cal D}\lambda_{a}{\cal
D}c^{a}{\cal D}
{\bar c}_{a}\exp\left[i\int dt{\widetilde{\cal L}}\right]
\end{equation} 

\noindent where $\widetilde{\cal L}$ is:

\begin{equation}
\label{eq:trentuno}
{\widetilde{\cal L}}=\lambda_{a}[{\dot\phi }^{a}-\omega^{ab}\partial_{b}H]+
i{\bar c}_{a}[\delta^{a}_{b}\partial_{t}-\omega^{ac}\partial_{c}\partial_{b}H]
c^{b}.
\end{equation} 

\noindent One can easily see that this Lagrangian gives the following equations of motion:
for $\phi$ and $c$:
\begin{eqnarray}
\label{eq:trentadue}
&&{\dot\phi }^{a}-\omega^{ab}\partial_{b}H = 0 \\
\label{eq:trentatre}
&&[\delta^{a}_{b}\partial_{t}-\omega^{ac}\partial_{c}\partial_{b}H]c^{b}=0. 
\end{eqnarray} 

\noindent One notices immediately the following two things:\\
\\
\noindent
{\bf 1)} $\widetilde{\cal L}$ leads to the same Hamiltonian equations for $\phi$ as
$H$ did;\\
\noindent
{\bf 2)} $c^b$ transforms under the Hamiltonian vector field $h\equiv\omega^{ab}
\partial_bH\partial_{a}$
as a {\it form} $d\phi^{b}$ does.\\
\\
From the above formalism, using some extended Poisson brackets ({\it EPB})
defined in the space \quattrova, one can get the equations of motion
also via an Hamiltonian $\widetilde{\cal H}$ given by

\begin{equation}
\label{eq:trentaquattro}
\widetilde{\cal
H}=\lambda_a\omega^{ab}\partial_bH+i\bar{c}_a\omega^{ac}(\partial_c\partial_bH)c^b.
\end{equation} 

\noindent The extended Poisson brackets mentioned above  are:
\begin{equation}
\label{trentacinque}
\{\phi ^{a},\lambda_{b}\}_{\scriptscriptstyle EPB}=\delta^{a}_{b};~~~~\{{\bar
c}_{b},
c^{a}\}_{\scriptscriptstyle EPB}=-i\delta^{a}_{b}. 
\end{equation}

\noindent The equations of motion are then given by 
$\frac{d}{dt}A=\{A,\widetilde{\cal H}\}_{\scriptscriptstyle EPB}$
where A is one of the variables \quattrova.
All the other {\it EPB} are zero; in particular $\{\phi^{a},\phi^{b}\}_
{\scriptscriptstyle EPB}=0$. This indicates that the
{\it EPB} are not the standard Poisson
brackets on ${\cal M}$ which would give $\{\phi^{a},\phi^{b}\}_
{\scriptscriptstyle PB}=\omega^{ab}$.

Being eq.(\ref{eq:trenta}) a path integral one could also introduce the concept of
{\it commutator}
as Feynman did in the quantum case. If we define the graded commutator of two
functions $O_1(t)$ and
$O_2(t)$ as the expectation value 
$\langle\ldots\rangle$ under our path integral of some time-splitting
combinations of the functions
themselves as:

\begin{equation}
\label{eq:trentasei}
\langle[O_{1}(t),O_{2}(t)]\rangle\equiv  \lim_{\epsilon\rightarrow 0}
\langle O_{1}(t+\epsilon)O_{2}(t)\pm O_{2}(t+\epsilon)O_{1}(t)\rangle, 
\end{equation}  

\noindent then we get  from eq.(\ref{eq:trenta}) that the only commutators different from
zero among the basic variables are:

\begin{equation}
\label{eq:trentasette}
\langle[\phi ^{a},\lambda_{b}]\rangle=i\delta^{a}_{b};~~~~\langle[{\bar
c}_{b},
c^{a}]\rangle=\delta^{a}_{b}.
\end{equation} 

\noindent We notice immediately two things:\\
\\
\noindent
{\bf A)} there is an isomorphism between the extended Poisson structure
(\ref{trentacinque}) and
the graded commutator structure (\ref{eq:trentasette}):
$\{.,.\}_{\scriptscriptstyle EPB}\longrightarrow -i[.,.]$;\\
\noindent
{\bf B)} via the commutator structure (\ref{eq:trentasette}) one can ``realize"
$\lambda_a$ and $\bar{c}_a$
as:

\begin{equation}
\label{eq:trentotto}
\lambda_{a}=-i{\partial\over\partial\phi ^{a}};~~~~{\bar
c}_{a}={\partial\over\partial c^{a}}.
\end{equation} 

\noindent It is now easy to check that, using eq.(\ref{eq:trentotto}), 
 what we got as weight
in eq.(\ref{eq:trenta})
corresponds to the operatorial version of CM. In fact take for the moment only
the bosonic part of $\widetilde{\cal H}$ in eq.(\ref{eq:trentaquattro}):~
${\widetilde{\cal H}}_{bos}=\lambda_{a}\omega^{ab}\partial_{b}H$; 
this one, via eq.(\ref{eq:trentotto}), goes into the operator\break
~${\widehat{\widetilde{\cal H}}}_{bos}\equiv
-i\omega^{ab}\partial_{b}H\partial_{a}$~
which is nothing else than the Liouville operator of CM. So we got
what we expected. If we had added the Grassmannian part to 
${\widetilde{\cal H}}_{bos}$ and inserted the operatorial representation (\ref{eq:trentotto}) of 
$\bar{c}$, we would have got an operator which
makes the evolution of functions depending not only on $\phi$ but also
on  $c$ like\break $F(\phi, c)=F_{a_{1}\ldots a_{p}}c^{a_{1}}\ldots c^{a_{2}}$.
Remembering that the $c^{a}$ transform as $d\phi^{a}$ (see point {\bf(1)} below
eq.(\ref{eq:trentatre})), we can say that the function $F(\phi,c)$ can be
put in correspondence with $p$-forms:

\be
\label{eq:trentanove}
F=F_{a_{1}\ldots a_{p}}c^{a_{1}}\ldots c^{a_{p}}\longrightarrow 
F_{a_{1}\ldots a_{p}}d\phi^{a_{1}}\wedge\ldots\wedge d\phi^{a_{p}}.
\ee

\noindent So our $\HT$ makes the evolution of forms that means it corresponds
to  the object known in literature\scite{MARS} as the Lie-Derivative of the 
Hamiltonian flow. Note  that we are not talking of forms built over the
space \quattrova  but only of forms over the space ${\mathcal M}$ 
whose coordinates are ($\phi^{a}$). Our $\HT$ is the Lie-derivative for this last space.
Via our variables it is also possible to build vector and multivector
fields over ${\mathcal M}$ and to reproduce the full Cartan calculus. 
For details we refer the
reader to ref.\scite{ENNIO} and for a deeper geometrical understanding
of our enlarged space~\quattrova~ we invite the interested reader
to consult refs.\scite{MARMAU}. 

The reader may remember that the concept
of Lie-derivative was mentioned also in the second of refs.\scite{WITTE}.
There anyhow the connection between Lie-derivative and Hamiltonian
was not as direct as here. Moreover the Lie-derivative was not associated
to the flow associated to the conformal potential but with the flow
associated to the superpotential (24).

The Hamiltonian $\HT$ has various {\it universal} symmetries\scite{ENNIO}
all of which have been studied geometrically. The associated charges\footnote{The charges which 
here have been indicated as  $C$ and ${\overline C}$ are those that in ref.\scite{ENNIO}
were called $K$ and ${\overline K}$. We change their names in order not
to confuse them with the $K$ operator of the conformal algebra
(see {\bf TABLE 1}).} are:

\begin{center}
{\bf TABLE 3}
\end{center}

\[
\begin{array}{|rcl|}
\hline &&\\
Q_{\scriptscriptstyle BRS} &=&\displaystyle i c^{a}\lambda_{a} \\

{\bar Q}_{\scriptscriptstyle BRS} &=&\displaystyle i {\bar
c}_{a}\omega^{ab}\lambda_{b} \\

Q_{g} &=&\displaystyle c^{a}{\bar c}_{a} \\

C &=&\displaystyle {\omega_{ab}c^{a}c^{b}\over 2} \\

{\bar C} &=&\displaystyle{\omega^{ab}{\bar c}_{a}{\bar c}_{b}\over 2} \\

\NH &=&\displaystyle c^{a}\partial_a H \\

{\overline \NH} &=&\displaystyle {\bar c}_{a}\omega^{ab}\partial_{b}H\\

&&\\
\hline
\end{array} 
\]

\noindent Using the correspondence between Grassmannian variables and forms,
the $\Qb$ turns out to be nothing else\scite{ENNIO} than the exterior 
derivative\footnote{We denoted some charges as BRS and anti-BRS charges ($\Qb$,$\QBb$) because 
they are 
exterior derivatives as the gauge-BRS charges are and because
they are nihilpotent and anticommutes among themselves. The $\omega_{ab}$
which appears in this table is the inverse of the $\omega^{ab}$ of eq.
(\ref{eq:venticinque}).} on phase space
and, as it is well known\scite{MARS} it always commutes with any
Lie-derivative. The $Q_{g}$, or ghost charge, is the form-number which 
is always conserved by the Lie-derivative. Similar geometrical meanings
can be found\scite{ENNIO} for the other charges that are listed above.
Of course linear combinations of them are also conserved and there are two
combinations which deserve our attention. They are the following charges:

\be
\label{eq:quaranta}
\QH\equiv\Qb-\beta\NH;~~~~~~~~~~\QBH\equiv\QBb+\beta{\overline\NH};
\ee

\noindent (where $\beta$ is a dimensionful parameter) 
which are true supersymmetry charges because, besides commuting
with $\HT$, they give\footnote{The commutators used below are those 
of our path-integral for classical mechanics derived in
eq.(\ref{eq:trentasette}).}:

\be
\label{eq:quarantuno}
[\QH,\QBH]=2i\beta\HT.
\ee

\noindent This proves that our $\HT$ is supersymmetric. To be precise it is an $N=2$
supersymmetry. One realizes immediately that $H$ acts
as a sort of superpotential for the supersymmetric Hamiltonian $\HT$. 
All this basically means that  we can obtain a  supersymmetric Hamiltonian $\HT$~out of any
system with Hamiltonian $H$~and, besides, our $\HT$ has a nice
geometrical meaning being the Lie-derivative of the Hamiltonian
flow generated by $H$. 

We will now build the $\HT$ of the conformal invariant system given
by the Hamiltonian of eq.(\ref{eq:otto}), that means we insert
the $H$ of eq.(\ref{eq:otto}) into the $\HT$ of eq.(\ref{eq:trentaquattro}).
The result is:
\be
\label{eq:quarantadue}
\HT=\lambda_q p+\lambda_p\frac{g}{q^3}
+i\bar{c}_qc^p-3i\bar{c}_pc^q\frac{g}{q^4}
\ee 
\noindent where the indices $(.)^{q}$ and $(.)^{p}$ on the variables $(\lambda, c, {\bar c})$ replace
the indices $(.)^{a}$ which appeared in the general formalism. 
In fact, as the system
is one-dimensional, the index ``$(.)^{a}$" can only indicate the variables ``$(p,q)$" and that is why
we use $(p,q)$ as index. The two supersymmetric 
charges of eq.(\ref{eq:quaranta}) are in this case
\bea
\label{eq:quarantatre}
&&\QH=\Qb+\beta\left(\frac{g}{q^3}c^q-pc^p\right) \\
\label{eq:qurantaquattro}
&&\QBH=\QBb+\beta\left(\frac{g}{q^3}\bar{c}_p+pc_q\right). 
\eea

\noindent It was one of the central points of the original paper\scite{DFF} on conformal
mechanics that the Hamiltonians of the system could be, beside $H_{0}$
of eq.(\ref{eq:sedici}), also $D_{0}$ or $K_{0}$ of eqs. (\ref{eq:diciassette})(\ref{eq:diciotto}) 
or any linear combination of them. In the same manner as we built the
Lie-derivative~$\HT$~associated to $H_{0}$, we can also
build the Lie-derivatives associated to the flow generated by $D_{0}$
and $K_{0}$. We just have to insert\footnote{We will neglect ordering
problem in the expression (\ref{eq:diciassette}) of $D_{0}$ because we are doing
a classical theory. The sub-index ``$(.)_{0}$" that we will put on $\DT$ and $\KT$
below is to indicate that they were built from the $D_{0}$ and $K_{0}$.}  $D_{0}$ or $K_{0}$ in place of $H$
as superpotential in the $\HT$ of eq.(\ref{eq:trentaquattro}). Calling the
associated Lie-derivatives as $\DT_0$ and 
$\KT_0$, what we get is:

\bea
\label{eq:quarantacinque}
&&\DT_0=\frac{1}{2}[\lambda_pp-\lambda_qq+i(\bar{c}_pc^p-\bar{c}_qc^q)]\\
\label{eq:quarantasei} 
&&\KT_0=-\lambda_pq-i\bar{c}_pc^q 
\eea

\noindent The construction is best illustrated in Figure 1.

\begin{figure}
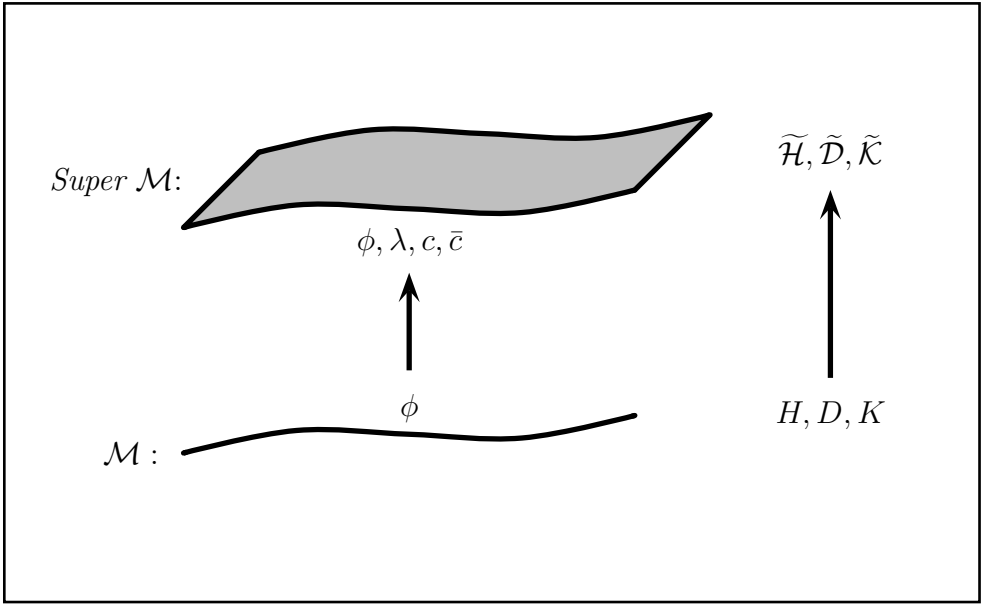

\pspicture(-1.25,0)(12.75,9)
\psset{linewidth=2pt}
\psframe[linewidth=1pt](0,0)(13,8)
\pscurve[]{c-c}(2.4,2)(3.9,2.3)(5.4,2.25)(6.9,2.2)(8.4,2.5)
\pscustom[fillstyle=solid,fillcolor=lightgray]{
\pscurve[]{c-c}(2.4,5)(3.9,5.3)(5.4,5.25)(6.9,5.2)(8.4,5.5)
\psline[](8.4,5.5)(9.4,6.5)
\pscurve[liftpen=2]{c-c}(9.4,6.5)(7.9,6.2)(6.4,6.25)(4.9,6.3)(3.4,6)
\psline[](3.4,6)(2.4,5)}

\rput(5.4,4.8){$\phi,\lambda,c,\bar{c}$}
\rput(5.4,2.6){$\phi$}
\psline[]{->}(5.4,3.1)(5.4,4.4)

\rput(11,6){$\HT,\DT,\KT$}
\rput(11,2.5){$H,D,K$}
\rput(1.7,2){$\mathcal M :$}
\rput(1.5,5.6){{\it Super} $\mathcal M$:}
\psline[]{->}(11,3)(11,5.5)
\endpspicture
\caption{\small The correspondence between $\mathcal M$ and $Super \mathcal M$.}
\end{figure}

As it is easy to prove, both $\DT_0$ and $\KT_0$ are supersymmetric. It is 
possible in fact to introduce the following charges:

\bea
\label{eq:quarantasette}
&&\QD=\Qb+\gamma(qc^p+pc^q);\\
\label{eq:quarantotto}
&&\QBD=\QBb+\gamma(p\bar{c}_p-q\bar{c}_q);\\
\label{eq:quarantanove}
&&\QK=\Qb-\alpha qc^q; \\
\label{eq:cinquanta}
&&\QBK=\QBb-\alpha q\bar{c}_p;
\eea

($\gamma$ and $\alpha$ play the same role as $\beta$ for $H$) which close on $\DT_0$ and $\KT_0$:

\bea
\label{eq:cinquantuno}
&&[\QD,\QBD]=4i\gamma\DT_0\\
\label{eq:cinquantadue}
&&[\QK,\QBK]=2i\alpha\KT_0.
\eea

\noindent One further point to notice is that 
the conformal algebra of eqs.(\ref{eq:undici})(\ref{eq:dodici})(\ref{eq:tredici}) 
is now realized, via the commutators 
(\ref{eq:trentasette}) of our formalism, by the $(\HT,\DT_0,\KT_0)$
and not by the old functions~$(H,D_0,K_0)$. In fact, via  these new commutators,
we get:

\bea
\label{eq:cinquantatre}
[\HT,\DT_0]=i\HT;~&~[\KT_0,\DT_0]=-i\KT_0;~&~[\HT,\KT_0]=2i\DT_0;\\
\label{eq:cinquantaquattro}
[H,D_0]=0;~&~[K_0,D_0]=0;~&~[H,K_0]=0.
\eea

\noindent The next thing to find out, assuming as basic Hamiltonian the $\HT$
and as supersymmetries the ones generated by $\QH$ and $\QBH$, is to perform
the commutators between supersymmetries and conformal operators so to 
get the superconformal generators. It is easy to work this out and we
get:

\bea
\label{eq:cinquantacinque}
[\QH,\DT_0]=i(\QH-\Qb); & &[\QBH,\DT_0]=i(\QBH-\QBb); \\
\label{eq:cinquantasei}
[\QH,\KT_0]=\frac{i\beta}{\gamma}(\QD-\Qb); & &[\QBH,\KT_0]=\frac{i\beta}{\gamma}(\QBD-\QBb).
\eea

\vskip .5cm
\noindent From what we have above we realize immediately the role of the $\QD$ and $\QBD$:
besides being the ``square roots" of $\DT_0$ they are also (combined with the
$\Qb$ and $\QBb$) the generators of the superconformal transformations.
It is also a simple calculation to evaluate the commutators
between the various ``supercharges" $\QH$,$\QBD$, $\QBK$, $\QBH$, $\QD$, $\QK$:

\bea
\label{eq:cinquantasette}
[\QH,\QBD]=i\beta\HT+2i\gamma\DT_0-2\beta\gamma H; & 
&[\QBH,\QD]=i\beta\HT+2i\gamma\DT_0+2\beta\gamma H; \\
\label{eq:cinquantotto}
[\QK,\QBD]=i\alpha\KT_0+2i\gamma\DT_0+2\alpha\gamma K_0; & 
&[\QBK,\QD]=i\alpha\KT_0+2i\gamma\DT_0-2\alpha\gamma K_0; \\
\label{eq:cinquantanove}
[\QH,\QBK]=i\beta\HT+\;i\alpha\KT_0-2\alpha\beta D_0; & 
&[\QBH,\QK]=i\beta\HT+i\alpha\KT_0+2\alpha\beta D_0;
\eea

\vskip .5cm
\noindent From the RHS of these expressions we see that one needs also the
old functions $(H,D_0,K_0)$ in order to close the algebra.

The complete set of operators which close the algebra is listed in the
following table:
\newpage
\begin{center}
{\bf TABLE 4}
\end{center}

\[
\begin{array}{|lcl|}
\hline & & \\
\HT=\displaystyle\lambda_q p+\lambda_p\frac{g}{q^3}+i\bar{c}_q
c^p-3i\bar{c}_pc^q\frac{g}{q^4}; & &H=\displaystyle\frac{1}{2}\left(p^2+\frac{g}{q^2}\right);\\
\KT_0=\displaystyle-\lambda_pq-i\bar{c}_pc^q; && K_0=\frac{1}{2}q^2;\\
\DT_0=\displaystyle\frac{1}{2}[\lambda_pp-\lambda_qq+i(\bar{c}_pc^p-\bar{c}_qc^q)];&&D_0=\displaystyle
-
\frac{1}{2}qp;\\
\Qb= \displaystyle i(\lambda_qc^q+\lambda_pc^p);&&\QBb= \displaystyle 
i(\lambda_p\bar{c}_q-\lambda_q\bar{c}_p);\\
\QH=\displaystyle\Qb+\beta\left(\frac{g}{q^3}c^q-pc^p\right);
&&\QBH=\displaystyle\QBb+\beta\left(\frac{g}{q^3}\bar{c}_p+p\bar{c}_q\right); \\
\QK=\displaystyle\Qb-\alpha qc^q; &&\QBK=\displaystyle\QBb-\alpha q\bar{c}_p;\\
\QD=\displaystyle\Qb+\gamma(qc^p+pc^q);&&\QBD=\displaystyle\QBb+\gamma(p\bar{c}_p-q\bar{c}_q);\\
&&\\
\hline
\end{array}
\]

The complete algebra among these generators is:

\begin{center}
{\bf TABLE 5}
\end{center}

\[
\hspace{-1.1cm}
\begin{array}{|lll|}

\hline 

& & \\

[\HT,\DT_0]=i\HT; \hspace{3cm}&[\KT,\DT_0]=-i\KT_0; \hspace{3cm}  &[\HT,\KT_0]=2i\DT_0; \\

[\QH,\HT]=0; &[\QBH,\HT]=0; &[\QH,\QBH]=2i\beta\HT; \\

[\QH,\DT_0]=i(\QH-\Qb); &[\QBH,\DT_0]=i(\QBH-\QBb); & \\

[\QH,\KT_0]=i\beta\gamma^{-1}(\QD-\Qb); &[\QBH,\KT_0]=i\beta\gamma^{-1}(\QBD-\QBb); & \\

[\Qb,\HT]=[\QBb,\HT]=0; & [\Qb,\KT]=[\QBb,\KT]=0; & [\Qb,\DT]=[\QBb,\DT]=0; \\

[\QD,\HT]=-2i\gamma\beta^{-1}(\QH-\Qb); &[\QBD,\HT]=-2i\gamma\beta^{-1}(\QBH-\QBb); & \\

[\QD,\KT_0]=2i\gamma\alpha^{-1}(\QK-\Qb); &[\QBD,\KT_0]=2i\gamma\alpha^{-1}(\QBK-\QBb); & \\

[\QD,\DT_0]=0; &[\QBD,\DT_0]=0; &[\QD,\QBD]=4i\gamma\DT_0; \\

[\QK,\HT]=-i\alpha\gamma^{-1}(\QD-\Qb); &[\QBK,\HT]=-i\alpha\gamma^{-1}(\QBD-\QBb); & \\

[\QK,\DT_0]=-i(\QK-\Qb); &[\QBK,\DT_0]=-i(\QBK-\QBb); & \\

[\QK,\KT_0]=0; &[\QBK,\KT_0]=0; &[\QK,\QBK]=2i\alpha\KT_0; \\

[\QH,\QBD]=i\beta\HT+2i\gamma\DT_0-2\beta\gamma H; & 
[\QBH,\QD]=i\beta\HT+2i\gamma\DT_0+2\beta\gamma H; & \\

[\QK,\QBD]=i\alpha\KT_0+2i\gamma\DT_0+2\alpha\gamma K; &
[\QBK,\QD]=i\alpha\KT_0+2i\gamma\DT_0-2\alpha\gamma K; & \\

[\QH,\QBK]=i\beta\HT+i\alpha\KT_0-2\alpha\beta D; & [\QBH,\QK]=i\beta\HT+i\alpha\KT_0+2\alpha\beta D; 
& \\

[\QH,\QBb]=[\QBH,\Qb]=i\beta\HT; & [\QK,\QBb]=[\QBK,\Qb]=i\alpha\KT_0; & \\

[\QD,\QBb]=[\QBD,\Qb]=2i\gamma\DT_0; & & \\

[Q_{\scriptscriptstyle(\ldots)},H]=\beta^{-1}(\Qb-\QH); & [\overline{Q}_{\scriptscriptstyle(\ldots)},H]= 
\beta^{-1}(\QBb-\QBH); & \\

[Q_{\scriptscriptstyle(\ldots)},D_0]=(2\gamma)^{-1}(\Qb-\QD); & 
[\overline{Q}_{\scriptscriptstyle(\ldots)},D_0]=(2\gamma)^{-1}(\QBb-\QBD); & \\

[Q_{\scriptscriptstyle(\ldots)},K_0]=\alpha^{-1}(\Qb-\QK); & 
[\overline{Q}_{\scriptscriptstyle(\ldots)},K_0]=\alpha^{-1}(\QBb-\QBK); & \\

[\HT,H]=0; & [\KT_0,K_0]=0; & [\DT_0,D_0]=0; \\

[\HT,K_0]=[H,\KT_0]=2iD; &
[\HT,D_0]=[H,\DT_0]=iH; & [\DT_0,K_0]=[D_0,\KT_0]=iK. \\

& & \\
\hline
\end{array}
\]

\noindent All other commutators\footnote{The $Q_{(\ldots)}$ appearing in the table can be
any of the following operators: $\Qb$,$\QH$,$\QD$,$\QK$ and the same holds for $\overline{Q}_{(\ldots)}$. 
Obviously 
all commutators are between quantities calculated at the same time.} are zero.

We notice than for our supersymmetric extension we need 14 charges
(see {\bf TABLE~4}) 
in order for the algebra to close, while in the extension of
ref.\scite{FUB} one needs only 8 charges (see {\bf TABLE 1}).
This is so not only because ours is an $N=2$ supersymmetry (while
the one of\scite{FUB} is an $N=1$) but also because of the totally
different character of the model.


\section{Study of the two Superconformal Algebras}

A Lie superalgebra\scite{KAC} is an algebra made of even $E_{n}$ and
odd $O_{\alpha}$ generators whose graded commutators look like:

\bea
\label{eq:sessanta}
[E_{m},E_{n}]&=&F^{p}_{mn}E_{p};\\
\label{eq:sessantuno}
[E_{m},O_{\alpha}]&=&G^{\beta}_{m\alpha}O_{\beta};\\
\label{Eq:sessantadue}
[O_{\alpha},O_{\beta}]&=&C^{m}_{\alpha\beta}E_{m};
\eea 

\noindent and where the structure constants $F^{p}_{mn},G^{\beta}_{m,\alpha},
C^{m}_{\alpha,\beta}$ satisfy generalized Jacobi identities.

One can interpret the relation (\ref{eq:sessantuno}) as saying that
the even part of the algebra has a representation on the odd part. 
This is clear if we consider the odd part as a vector space and that
the even part acts on this vector space via the graded
commutators. The structure constants $F^{\beta}_{m\alpha}$ are then
the matrix elements which characterize the representations.

For superconformal algebras the usual folklore says that the even part of the
algebra has his conformal subalgebra represented spinorially
on the odd part. The reasoning roughly
goes as follows: the odd part of the algebra must contain the supersymmetry
generators which transform as spinors under the Lorentz group which is
a subgroup of the conformal algebra. So it is impossible that  the whole
conformal algebra  is represented non-spinorially on the odd part.

Actually this line of reasoning is true in a relativistic context in which 
the supersymmetry is a true relativistic supersymmetry and the charges must 
carry a spinor index due to their nature. In our non-relativistic
point particle case  instead the charges  do not carry any space-time index 
and so we do not have as a consequence that necessarily the even part of the
algebra is represented spinorially on the odd part. It can happen but it can
also not happen. In this respect we will analyze the superalgebras
of the two supersymmetric extensions of conformal mechanics seen here, the
one of\scite{FUB} and ours.

Let us start from the one of ref.\scite{FUB}  which is given in {\bf TABLE 1}.
The conformal subalgebra ${\mathcal G}_{0}$ of the even part  can be 
organized in an $SO(2,1)$ form as follows:

\[
{\mathcal G}_{0}: \left\{ 
\begin{array}{l}
B_{1}=\displaystyle{1\over 2}\left[ {K\over a}-aH \right] \\
B_{2}= D \\
J_{3}=\displaystyle{1\over 2}\left[ {K\over a}+a H \right]
\end{array}
\right.
\]

\noindent where $a$ is the same parameter introduced in \cite{DFF} with dimension of
time.

The odd part ${\mathcal G}_{1}$ is:

\[
{\mathcal G}_{1}: \left\{ \begin{array}{l}
Q\\
Q^{\dag}\\
S\\
S^{\dag}
\end{array}
\right.
\]

\noindent It is easy to work out, using the results of {\bf TABLE 2}, the action
of the ${\mathcal G}_{0}$  on ${\mathcal G}_{1}$. The result
is summarized in the following table:

\begin{center}
{\bf TABLE 6}
\end{center}

\[
\begin{array}{|lll|}
\hline & & \\

[B_{1},Q]=\displaystyle{1\over 2a}S; \hspace{2cm}&[B_{2},Q]=-\displaystyle{i\over 2}Q; \hspace{2cm} &
\displaystyle[J_{3},Q]={1\over 2a}S; \\

[B_{1},Q^{\dag}]=\displaystyle-{1\over 2a}S^{\dag}; \hspace{2cm}&[B_{2},Q^{\dag}]=
\displaystyle-{i\over 2}Q^{\dag}; \hspace{2cm} & [J_{3},Q^{\dag}]=\displaystyle-{1\over 2a}S^{\dag};\\

[B_{1},S]=-\displaystyle{a\over 2}Q; \hspace{2cm}&[B_{2},S]=\displaystyle{i\over 2}S; \hspace{2cm} &
[J_{3},S]=\displaystyle{a\over 2}Q; \\

[B_{1},S^{\dag}]=\displaystyle{a\over 2}Q^{\dag}; \hspace{2cm}&[B_{2},S^{\dag}]=\displaystyle{i\over 2}
S^{\dag}; \hspace{2cm} & [J_{3},S^{\dag}]=\displaystyle-{a\over 2}Q^{\dag}; \\

&& \\
\hline
\end{array}
\]

\noindent As we said before, in order to act with the even part of the
algebra on the odd part, we have to consider the odd part of the as a vector
space. Let us then introduce the following ``vectors":

\bea
\label{eq:sessantatre}
|q\rangle&\equiv & Q+Q^{\dag}\\
\label{eq:sessantaquattro}
|p\rangle & \equiv & S-S^{\dag}\\
\label{eq:sessantacinque}
|r\rangle & \equiv & Q-Q^{\dag}\\
\label{eq:sessantasei}
|s\rangle & \equiv & S+S^{\dag};
\eea

\noindent they label a 4-dimensional vector space.
On these vectors we act via the commutators, for example:

\be
\label{eq:sessantasette}
B_{1}|q\rangle\equiv [B_{1},Q+Q^{\dag}]
\ee

\noindent It is then immediate  to realize from {\bf TABLE 6} that the 2-dim. space with
basis ~$(|q\rangle,|p\rangle)$ form a closed space under the action
of even part of the algebra so it carries a 2-dim. representation
and the same holds for the space $(|r\rangle,|s\rangle)$. We can immediately check
which kind of representation is this: Let us take the Casimir operator
of the algebra ${\mathcal G}_{0}$ which is~
${\mathcal C}=B_{1}^{2}+B_{2}^{2}-J_{3}^{2}$
and apply it to a state of one of the two 2-dim. representations:

\bea
\label{eq:sessantotto}
{\mathcal C}|q\rangle & = & [B_{1},[B_{1},Q+Q^{\dag}]]+[B_{2},[B_{2},Q+Q^{\dag}]]-
[J_{3},[J_{3},Q+Q^{\dag}]]\nonumber \\
& = & -{3\over 4}(Q+Q^{\dag})\nonumber\\
& = & -{3\over 4}|q\rangle
\eea

\noindent This factor ${3\over 4}=-{1\over 2}({1\over 2}+1)$ indicates that the
$(|q\rangle,|p\rangle)$ space carries a spinorial representation. It is possible to prove the same for the
other space.

Let us now turn the same crank for our supersymmetric extension
of conformal mechanics. Looking at the {\bf TABLE 4} of our operators,
we can organize the even part ${\mathcal G}_{0}$, as follows:

\vskip 1cm
\begin{center}
{\bf TABLE 7 (${\mathcal G}_{0}$)}
\end{center}
\[
\begin{array}{|lcl|}
\hline 

&&\\
B_{1}=\displaystyle{1\over 2}\left( {\KT\over a}-a\HT\right); && P_{1}=2D;\\

B_{2}=\displaystyle\DT; && P_{2}=\displaystyle aH-{K\over a};\\

J_{3}=\displaystyle{1\over 2}\left( {\KT\over a}+a\HT\right); && P_{0}=\displaystyle aH+{K\over a};\\
&&\\
\hline
\end{array}
\]
\vskip 1 cm
\noindent The LHS is the usual $SO(2,1)$ while the RHS is formed by three translations
because they commute among themselves. So the overall algebra is the Euclidean
group $E(2,1)$.

The odd part of our superalgebra is made of 8 operators (see {\bf TABLE 4})
which are:
\newpage
\begin{center}
{\bf TABLE 8 (${\mathcal G}_{1}$)}
\end{center}
\[
\begin{array}{|lcl|}
\hline & & \\

\QH; && \QBH;\\

\QK; && \QBK;\\

\QD; && \QBD;\\

\Qb; && \QBb;\\
& & \\
\hline
\end{array}
\]

\noindent As we did before in {\bf TABLE 6} for the model of\scite{FUB},
we will now evaluate for our model the action of ${\mathcal G}_{0}$ 
on ${\mathcal G}_{1}$. The result is summarized in the next table:

\begin{center}
{\bf TABLE 9}
\end{center}

\[
\begin{array}{|ll|}
\hline
& \\

[B_1,\QH]=\displaystyle\frac{i}{2\eta}(\Qb-\QD); & [B_1,\QBH]=\displaystyle\frac{i}{2\eta}(\QBb-\QBD);\\

[B_1,\QK]=\displaystyle\frac{i}{2\eta}(\Qb-\QD); & [B_1,\QBK]=\displaystyle\frac{i}{2\eta}(\QBb-\QBD);\\

[B_1,\QD]=-i\eta(\QH+\QK-2\Qb); &
[B_1,\QBD]=-i\eta(\QBH+\QBK-2\QBb);\\

[B_1,\Qb]=0; & [B_1,\QBb]=0;\\

[B_2,\QH]=i(\Qb-\QH); & [B_2,\QBH]=i(\QBb-\QBH);\\

[B_2,\QK]=i(\QK-\Qb); & [B_2,\QBK]=i(\QBK-\QBb);\\

[B_2,\QD]=0; & [B_2,\QBD]=0;\\

[B_2,\Qb]=0; & [B_2,\QBb]=0;\\

[J_3,\QH]=\displaystyle\frac{i}{2\eta}(\Qb-\QD); & [J_3,\QBH]=\displaystyle\frac{i}{2\eta}(\QBb-\QBD);\\

[J_3,\QK]=\displaystyle -\frac{i}{2\eta}(\Qb-\QD); & [J_3,\QBK]=\displaystyle -\frac{i}{2\eta}(\QBb-\QBD);\\

[J_3,\QD]=i\eta(\QH-\QK); & [J_3,\QBH]=i\eta(\QBH-\QBK);\\

[J_3,\Qb]=0; & [J_3,\QBb]=0;\\

[P_1,Q_{\scriptscriptstyle(\ldots)}]=\gamma^{-1}(\QD-\Qb); & 
[P_1,\overline{Q}_{\scriptscriptstyle(\ldots)}]=-\gamma^{-1}(\QBD-\QBb);\\

[P_2,Q_{\scriptscriptstyle(\ldots)}]=\gamma^{-1}\eta(\QH-\QK); & 
[P_2,\overline{Q}_{\scriptscriptstyle(\ldots)}]=-\gamma^{-1}\eta(\QBH-\QBK);\\

[P_0,Q_{\scriptscriptstyle(\ldots)}]=\gamma^{-1}\eta(\QH+\QK-2\Qb); & 
[P_0,\overline{Q}_{\scriptscriptstyle(\ldots)}]=-\gamma^{-1}\eta(\QBH+\QBK-2\QBb);\\

& \\

\hline
\end{array}
\]

\noindent where for simplicity we have made the choice $\displaystyle a=\sqrt{\frac{\beta}{\alpha}}$ 
and $\displaystyle\eta\equiv\frac{\gamma}{\sqrt{\alpha\beta}}$.

As we have to represent the conformal subalgebra of ${\mathcal G}_{0}$
(see {\bf TABLE 7}) on the vector space ${\mathcal G}_{1}$ of {\bf TABLE 8} 
it is easy to realize from {\bf TABLE 9} that the following three vectors

\be
\left\{
\begin{array}{l}
|q_{\scriptscriptstyle H}\rangle = (\QH-\Qb)-(\QBH-\QBb)\\
|q_{\scriptscriptstyle K}\rangle = (\QK-\Qb)-(\QBK-\QBb)\\
|q_{\scriptscriptstyle D}\rangle = \eta^{-1}[(\QD-\Qb)-(\QBD-\QBb)]
\end{array}
\right.
\ee
\noindent
make an irreducible representation of the conformal subalgebra. In fact, using\break
{\bf TABLE~9}, we get:

\be
\left\{ 
\begin{array}{l}
B_{1}|q_{\scriptscriptstyle H}\rangle =-{i\over 2}
|q_{\scriptscriptstyle D}\rangle\\
B_{2}|q_{\scriptscriptstyle H}\rangle=-i|q_{\scriptscriptstyle H}\rangle\\
J_{3}|q_{\scriptscriptstyle H}\rangle=-{i\over 2}
|q_{\scriptscriptstyle D}\rangle\\
B_{1}|q_{\scriptscriptstyle K}\rangle=-{i\over 2}
|q_{\scriptscriptstyle D}\rangle\\
B_{2}|q_{\scriptscriptstyle K}\rangle=i|q_{\scriptscriptstyle K}\rangle\\
J_{3}|q_{\scriptscriptstyle K}\rangle={i\over 2}|q_{\scriptscriptstyle
D}\rangle\\
B_{1}|q_{\scriptscriptstyle D}\rangle= -i(|q_{\scriptscriptstyle H}\rangle
+|q_{\scriptscriptstyle K}\rangle)\\
B_{2}|q_{\scriptscriptstyle D}\rangle=0\\
J_{3}|q_{\scriptscriptstyle D}\rangle=i(|q_{\scriptscriptstyle H}\rangle-
|q_{\scriptscriptstyle K}\rangle).
\end{array}
\right.
\ee

\noindent Having three vectors in this representation we presume it is an `integer'
spin representation, but to be sure let us apply the Casimir operator
on a vector. The Casimir is given, as before, by: 
${\mathcal C}=B_{1}^{2}+B_{2}^{2}-J_{3}^{2}$ but we must remember to use as 
$B_1$, $B_2$ and $J_3$ the operators contained
in {\bf TABLE~7}. It is then easy to check that
\be
\label{eq:settantuno}
{\mathcal C}|q_{\scriptscriptstyle H}\rangle=-2|q_{\scriptscriptstyle H}\rangle.
\ee
The same we get for the other two vectors $|q_{\scriptscriptstyle K}\rangle,
|q_{\scriptscriptstyle D}\rangle$, so the eigenvalue in the equation above is $-2=-1(1+1)$ 
and this indicates that those vectors make a ``spin" 1 representation.

In the same way as before it is easy to prove that these other three vectors:
\be
\left\{
\begin{array}{l}
|\widetilde{q_{\scriptscriptstyle H}}\rangle= (\QH-\Qb)+(\QBH-\QBb)\\
|\widetilde{q_{\scriptscriptstyle K}}\rangle= (\QK-\Qb)+(\QBK-\QBb)\\
|\widetilde{q_{\scriptscriptstyle D}}\rangle=(\QD-\Qb)+(\QBD-\QBb)
\end{array}
\right.
\ee
\noindent
make another irreducible representation of ``spin" 1.

Of course, as the vector space ${\mathcal G}_{1}$ of {\bf TABLE 8} is
8-dimensional and up to now we have used  only 6 vectors  to build the
two integer representations, we expect that there must be
some other representations which can be built using the two remaining vectors. It is in fact so.
We can build the following two other vectors:
\bea
\label{eq:settantatre}
|q_{\scriptscriptstyle BRS}\rangle=\Qb-\QBb\\
\label{eq:settantaquattro}
|\widetilde{q_{\scriptscriptstyle BRS}}\rangle=\Qb+\QBb
\eea
\noindent
and it is easy to see that each of them carry a representation of spin zero:
\be
\label{eq:settantacinque}
{\mathcal C}|q_{\scriptscriptstyle BRS}\rangle={\mathcal C}|\widetilde{q_{\scriptscriptstyle BRS}}\rangle=0
\ee 
So we can conclude that our vector space ${\mathcal G}_{1}$
carries a reducible representation of the conformal algebra
made of two spin one and two spin zero representations.

We wanted to do this analysis in order to underline a further difference
between our supersymmetric extension and the one of\scite{FUB}
whose odd part ${\mathcal G}_{1}$, as we showed before, carries
two spin one-half representations.

One last thing to do is to find out to which of the superalgebras classified in 
the literature ours belongs. We will come back to this in the future.


\section{Superspace Formulation of the Model}

\hspace{.4 cm}It is easy and instructive to do a superspace formulation of our
model like the authors of ref.\cite{FUB} did for theirs.

Let us enlarge our ``base space" $(t)$ to a superspace $(t,\theta,{\bar\theta})$
where $(\theta, {\bar\theta})$ are Grassmannian partners of $(t)$.
It is then possible to put all the variables \quattrova~in a single superfield
$\Phi$ defined as follows:

\be
\label{eq:settantasei}
\Phi^{a}(t,\theta,{\bar \theta})=\phi^{a}(t)+\theta c^{a}(t)+
{\bar \theta}~\omega^{ab}{\bar c}_{b}(t)+i{\bar\theta}\theta~\omega^{ab}\lambda_{b}(t)
\ee

This superfield had already been introduced in ref.\cite{ENNIO}. It is
a scalar field under the supersymmetry transformations of the system.
The various factors of ``$i$" appearing in its definition are due to the
fact that we chose \cite{ENNIO} the $c^{a},{\bar c}_a$ to be real and
the $\theta,{\bar\theta}$ to be pure imaginary.

It is a simple exercise to find the expansion of any function $F(\Phi^{a})$ 
of the superfields in terms of $\theta,{\bar\theta}$. For example, 
choosing as function the Hamiltonian $H$ of a system, we get:

\be
\label{eq:settantasette}
H(\Phi^{a})=H(\phi)+\theta N_{\scriptscriptstyle H}-{\bar\theta}~
{\overline N}_{\scriptscriptstyle H}+i\theta{\bar\theta}~\HT
\ee

\noindent
where  $\NH$ and ${\overline \NH}$ and $\HT$ are those given in
{\bf TABLE 3} and in eq.(\ref{eq:trentaquattro}).

From eq.(\ref{eq:settantasette}) it is easy to prove that:

\be
\label{eq:settantotto}
i\int H(\Phi)~d\theta d{\bar\theta}=\HT
\ee

\noindent Here we immediately notice a crucial difference with the supersymmetric
QM model of ref.\cite{FUB}. In the language of superfields (see the second of ref.\scite{FUB})
those authors obtain the supersymmetric potential
of their Hamiltonian  by inserting
the superfield into the superpotential (which is given by
eq.({\ref{eq:ventiquattro})) and integrating in  something
like $\theta,{\bar\theta}$, while we get the potential part
of our supersymmetric Hamiltonian by inserting the superfield into the
normal potential of the conformal mechanical model given in (\ref{eq:sedici}).

The space \quattrova  somehow can be considered 
as a {\it target} space whose {\it base} space is the superspace 
$(t,\theta,{\bar\theta})$. The action of the various charges listed in our\break
{\bf TABLE~4} is on the target-space variables but we can consider it as
induced by some transformations on the base-space. If we collectively
indicate  the charges acting on \quattrova  as $\Omega$, we shall indicate
the generators of the corresponding transformations on the base space
as ${\widehat \Omega}$. The relation between the two is the following:

\be
\label{eq:settantanove}
\delta\Phi^{a}=-\varepsilon {\widehat \Omega}\Phi^{a}
\ee
where 
\be
\label{eq:ottanta}
\delta\Phi^{a}=[\varepsilon\Omega,\Phi^{a}]
\ee
with $\varepsilon$ the commuting or anticommuting infinitesimal parameter
of our transformations\footnote{The conventions (\ref{eq:settantanove}) and
(\ref{eq:ottanta}) are slightly different than the ones in ref.\scite{ENNIO}.
Here we also correct some misprints present in that reference.} and $[(.),(.)]$
the graded commutators of eq.(\ref{eq:trentasette}).

Using the relations above it is easy to work out the superspace representation
of the operators contained in {\bf TABLE 3}, they are given in the table below:

\begin{center}
{\bf TABLE 10}
\end{center}

\[
\begin{array}{|rcl|}
\hline &&\\
\QCb &=& -\partial_{\theta} \\

\QCBb &=& \partial_{\bar\theta} \\

{\widehat Q}_{g} &=& {\bar\theta}\partial_{\bar\theta}-
\theta\partial_{\theta}  \\

{\widehat C} &=& {\bar\theta}\partial_{\theta} \\

{\widehat {\overline C}} &=& \theta\partial_{\bar\theta}\\

\widehat{N}_H &=& {\bar\theta}\partial_{t}\\

\widehat{\overline{N}}_H &=& \theta\partial_{t}\\

&&\\
\hline
\end{array} 
\]
\vskip .5cm
\noindent Via the charges above it is immediate to write down also the
supersymmetric charges of eq.(\ref{eq:quaranta}):

\be
\label{eq:ottantuno}
\QCH=-{\partial}_{\theta}-\beta~{\bar\theta}\partial_{t};~~~~~~~~~~
\QCBH={\partial}_{\bar\theta}+\beta~{\theta}\partial_{t};
\ee
Their anticommutator gives:
\be
\label{eq:ottantadue}
[\QCH,\QCBH]=-2\beta {\partial\over\partial t}
\ee
\noindent
from which, comparing the above equation with eq.(\ref{eq:quarantuno}),
one gets the superspace representation of $\HT$:
\be
\label{eq:ottantatre}
\HCT=i{\partial\over\partial t}.
\ee

\noindent Proceeding in the same way, via the relations (\ref{eq:settantanove}),
(\ref{eq:ottanta}), it is a long but easy procedure to give a superspace
representation to the charges $\QD,\QBD,\QK,\QBK$ of
eqs. (\ref{eq:quarantasette})-(\ref{eq:cinquanta}). This long derivation 
is contained in the appendix and the result is:

\bea
\label{eq:ottantaquattro}
\QCK &=& -{\partial\over\partial\theta}-\alpha~\omega^{ad}K_{db}~{\bar\theta}
\\
\label{eq:ottantacinque}
\QCBK &=& {\partial\over\partial{\bar\theta}}+\alpha~\omega^{ad}K_{db}~\theta\\
\label{eq:ottantasei}
\QCD & = & -{\partial\over\partial\theta}-2\gamma~\omega^{ad}D_{db}~{\bar\theta}\\
\label{eq:ottantasette}
\QCBD & = & {\partial\over\partial{\bar\theta}}+2\gamma~\omega^{ad}D_{db}~\theta
\eea
\noindent
where the matrices $K_{db}$ and $D_{db}$ are:
\be
\label{eq:ottantotto}
K_{db}  = 
\left(\begin{array}{lr}
1 & 0\\
0 & 0
\end{array}
\right);~~~~~~~
D_{db}  = 
-{1\over 2}\left(\begin{array}{lr}
0 & 1\\
1 & 0
\end{array}
\right)
\ee
(the repeated indices in eqs.(\ref{eq:ottantaquattro})-(\ref{eq:ottantasette})
are summed). The matrices $K_{db}$ and $D_{db}$ are $2\times 2$ just because
the symplectic matrix itself $\omega^{ab}$ is $2\times 2$ in our case. The
conformal mechanics system in fact has just a pair of phase-space
variables $(p,q)$ and the index in the $\phi^{a}$-phase space variables
can take only 2 values to indicate either $q$ or $p$ (see the eqs. of motion
(\ref{eq:venticinque})).

From the expressions of $\QCK,\QCBK,\QCD,\QCBD$ above we see that they have
two free indices. This implies (see eq.(\ref{eq:settantanove})) that those
operators ``turn" the various superfields in the sense that they turn a
$\Phi^{q}$~into combinations of $\Phi^{q}$ and $\Phi^{p}$ and viceversa.
This is something the other charges did not do.

Reached this point, we should  stop and think a little
bit about this superspace representation. We gave the superspace
representation of the various charges\break
$(\QD,\QBD,\QK,\QBK)$ of eqs.(47)-(50) which were linked to the
$\DT_0,\KT_0$ of eqs.(45)(46). But these last quantities were built
using the $D_{0}$ and $K_{0}$ that is the $D$ and $K$ at $t=0$.
If we had used, in building the $\DT_0, \KT_0$, the $D$ and $K$ at $t\neq 0$
of eqs.(9) and (10), we would have obtained a $\DT$ and a $\KT$ different
from those of eqs.(45)(46) and which would have had an explicit
dependence on $t$. Consequently also the associated supersymmetric
charges $(Q^t_{\scriptscriptstyle D},\overline{Q}^t_{\scriptscriptstyle D},
Q^t_{\scriptscriptstyle K},\overline{Q}^t_{\scriptscriptstyle K})$, having extra terms depending on $t$, 
would be different from those of eqs.(47)-(50). Being these charges different,
also their superspace representations shall be different from those 
given in eqs.(84)-(87). The difference at the level of superspace is crucial
because it involves $t$ which is part of superspace.

Let us then start this over-all process by first building the explicitly $t$-dependent
$\DT$ and $\KT$ from the following operators $D$ and $K$:

\bea
\label{eq:ottantanove}
H &= & H_{0}\\
\label{eq:novanta}
D & = & t H+D_{0}\\
\label{eq:novantuno}
K & = & t^{2}H+2tD_{0}+K_{0}
\eea

\noindent

from which we get:

\bea
\label{eq:novantadue}
\DT & = & t\HT+\DT_{0}\\
\label{eq:novantatre}
\KT & = & t^{2}\HT+2t\DT_{0}+\KT_{0}.
\eea
\vskip .5cm

\noindent It is easy to understand why these relations hold by remembering the
manner we got the Lie-derivatives out of the superpotentials.
The explicit form of $\DT$ in terms of \quattrova
can be obtained from (92) once we insert the $\HT$ and $\DT_{0}$
whose explicit form we already had in eqs.(42) and (45). The same for
$\KT$. Let us now turn to the form of the
associated fermionic charges which we will indicate as
$(\QDT,\QKT,\QBDT,\QBKT)$ where the index ``$(.)^{t}$" is to indicate
their explicit dependence on $t$. As it is shown in formula (A.1) of 
Appendix A, the $\QD$ and $\QK$ can be written using the charges $\ND$
and $\NK$ of (A.2) and the $\Qb$. As it is only the $N_{(\ldots)}$ and not the
$\Qb$ which pull in quantities like $D,K$ which may depend 
explicitly on time, we should only concentrate on the $N_{(\ldots)}$.
From their definition (see eq.(A.2)):

\be
\label{eq:novantaquattro}
\ND=c^{a}\partial_{a}D;~~~~~~~
\NK=c^{a}\partial_{a}K
\ee

\noindent we see that applying the operator $c^{a}\partial_{a}$ on both sides 
of eqs.(90)(91), we get:

\bea
\label{eq:novantacinque}
\NDT & = & t\NH+\ND\\
\label{eq:novantasei}
\NKT & = & t^{2}\NH+2t\ND+\NK.
\eea
\vskip .5cm

\noindent The next step is to write the $\QDT$ and $\QKT$. As they are given in
formula (A.1), using that equation and (95)(96) above
we get:

\bea
\label{eq:novantasette}
\QDT & = &\Qb-2\gamma N^t_{\scriptscriptstyle D}\\
\label{eq:novantotto}
\QKT & = & \Qb-\alpha N^t_{\scriptscriptstyle K}.
\eea
\vskip 1cm

\noindent In a similar manner, via eq.(A.9) and applying the operator
${\bar c}_{a}\omega^{ab}\partial_{b}$ to eqs.(90)(91), we get the $\QBDT$
and $\QBKT$:

\bea
\label{eq:novantanove}
\QBDT & = & \QBb+2\gamma\overline{N}^t_{\scriptscriptstyle D}\\
\label{eq:cento}
\QBKT & = & \QBb+\alpha\overline{N}^t_{\scriptscriptstyle K}.
\eea

\noindent We shall not write down explicitly the expressions of $(\QDT,\QKT,\QBDT,\QBKT)$
in terms of \quattrova because we have already 
in eqs.(47)-(50) and (A.2)(A.9) the expressions\footnote{Only be careful
in using the $D_{0}$ and $K_{0}$ in eqs.(A.1) and (A.9).} of the various charges
$(\QD,\QK,\QBD,\QBK,\ND,\NK, {\overline\NK},{\overline\ND})$ which make up,
according to eqs.(97)-(100), the new time dependent charges.
The next step is to obtain the superspace version of $(\QDT,\QKT,\QBDT,\QBKT)$.
Following a procedure identical to the one explained in detailed in the
appendix for the time-independent charges it is easy to get them and, via
their anticommutators, to derive the superspace version of the
$\DT$ and $\KT$. All these
operators are listed in the table below:

\begin{center}
{\bf TABLE 11}
\end{center}

\[
\begin{array}{|rcl|}
\hline && \\
\HCT&=&\displaystyle i\frac{\partial}{\partial t};\\
\DCT&=&\displaystyle i t\frac{\partial}{\partial t}-\frac{i}{2}\sigma_3;\\
\KCT&=&\displaystyle it^2\frac{\partial}{\partial t}-it\sigma_3-i\sigma_-;\\
\widehat{Q}^t_{\scriptscriptstyle D}&=&\displaystyle-\frac{\partial}{\partial\theta}-
2\gamma~{\bar\theta}t\frac{\partial}{\partial
t}+\gamma~{\bar\theta}\sigma_3;\\
\widehat{\overline{Q}}^t_{\scriptscriptstyle 
D}&=&\displaystyle\frac{\partial}{\partial\bar{\theta}}+2\gamma~\theta
t\frac{\partial}{\partial t}-\gamma{\theta}\sigma_3;\\
\widehat{Q}^t_{\scriptscriptstyle 
K}&=&\displaystyle-\frac{\partial}{\partial\theta}-\alpha~{\bar\theta}t^2\frac{\partial}{\partial
t}+\alpha~t{\bar\theta}\sigma_3+\alpha~{\bar\theta}\sigma_{-};\\
\widehat{\overline{Q}}^t_{\scriptscriptstyle K}&=&\displaystyle\frac{\partial}{\partial\bar{\theta}}+\alpha\theta~
t^2\frac{\partial}{\partial t}-\alpha t~\theta\sigma_3-\alpha\theta\sigma_{-}; \\
\hat{H}&=&\displaystyle\bar{\theta}\theta\frac{\partial}{\partial t};\\
\hat{D}&=&\displaystyle\bar{\theta}\theta~(t\frac{\partial}{\partial
t}-\frac{1}{2}\sigma_3);\\
\hat{K}&=&\displaystyle\bar{\theta}\theta~(t^2\frac{\partial}{\partial
t}-t\sigma_3-\sigma_{-}).\\
&& \\
\hline
\end{array}
\]

\noindent In the previous table the $\sigma_{3}$ and $\sigma_{-}$ are the Pauli
matrices:

\be
\label{eq:centouno}
\sigma_{3}  = 
\left(\begin{array}{lr}
1 & 0\\
0 & -1
\end{array}
\right);~~~~~~~
\sigma_{-}  = 
\left(\begin{array}{lr}
0 & 0\\
1 & 0
\end{array}
\right).
\ee

\noindent The reasons for the presence of these two-dimensional matrices
has been explained in the paragraph below eq.(88).

The last three operators listed in {\bf TABLE 11} are the
superspace version of the old $(H,D,K)$. To get this representation
we used again and again the rules given by eqs.(79)(80). As their representation
looks quite unusual, we have reported the details of their
derivations in Appendix B.

We want to conclude this section by presenting a new symmetry of our
system. A symmetry which is not among those found up-to-now whose
charges we have listed in {\bf TABLE 4}. It is associated with the following
superspace operator:

\be
\label{eq:centodue}
{\widehat{Q}_S}=\theta{\partial\over\partial\theta}+{\bar\theta}{\partial
\over{\partial\bar\theta}}.
\ee

\noindent This operator  is very similar to the ghost-charge ${\hat Q}_{\scriptscriptstyle g}$
of {\bf TABLE 10} but it has a crucial sign difference.

Let us apply it on the RHS of eq.(79) and see which variation it induces
on the \quattrova:

\be
\label{eq:centotre}
\delta_{{\scriptscriptstyle \QS}}\Phi^{a} =-\varepsilon{\widehat{Q}_S}\Phi^{a}=
-\varepsilon(\theta c^{a}+{\bar\theta}\omega^{ab}{\bar c}_{b}+2i{\bar\theta}
\theta\omega^{ab}\lambda_{b}).
\ee

\noindent Comparing the components with the same number of $\theta$ and ${\bar\theta}$
on both side of the equation above, we get 

\bea
\label{eq:centoquattro}
\delta_{{\scriptscriptstyle\QS}}\phi^{a} & = & 0\\
\delta_{{\scriptscriptstyle\QS}}c^{a} & = & -\varepsilon c^{a}\\
\delta_{{\scriptscriptstyle\QS}}{\bar c}_{a} & = &-\varepsilon {\bar c}_{a}\\
\delta_{{\scriptscriptstyle\QS}}\lambda_{a} & = & -2\varepsilon \lambda_{a}.
\eea 

\noindent It is easy to check how the Lagrangian of our model (see eq.(31)) changes
under the variations above:

\be
\label{eq:centootto}
\delta_{\QS}{\LT}= -2\varepsilon\LT.
\ee

\noindent
We can conclude that the transformations induced by $\widehat{Q}_S$
on the \quattrova are a symmetry of our system. In fact they just rescale
the overall lagrangian so they keep the equations of motion invariant
and these qualifies them as symmetry transformations. Of course
they are {\it non-canonical} symmetries because the rescaling of the whole
lagrangian is not a canonical transformation in \quattrova.
As it is not canonical we cannot find a canonical generator 
in \quattrova associated to $\widehat{Q}_S$. Note that the above
is a symmetry of any $\LT$ not necessarily of the conformal model
we have analyzed here. We could ask if this was a symmetry also of
the susyQM model of Witten\scite{WITTE} or at least of the
conformal QM of ref.\scite{FUB}. The answer is {\it No!}, the 
technical reason
being that in those QM models the analogs of the $\lambda_{a}$ variables
enter the lagrangian with a quadratic term while in our $\LT$ they enter linearly.
There is also an important physical reason why that symmetry was not present in those
 QM models while it was present in our CM one. 
The reason is that in QM one cannot rescale the action (as our symmetry
does) because there is the $\hbar$ setting a scale for the action, while it
can be done in CM where no scale is set. The reader may object that our
transformation rescales the Lagrangian but not the action

\be
{\tilde S}=\int \LT~dt 
\ee

\noindent because one could compensate the rescaling of the $\LT$ with a rescaling
of $t$. That is not so because our $\QS$ transforms only the Grassmannian
partners of time ($\theta,{\bar\theta})$ and not time itself.

We will come back to this symmetry in the future because it seems to
be at the heart of the difference between QM and CM.

\section{Exact Solution of the Supersymmetric Model} 

The original conformal mechanical model was solved exactly
in eq.(2.35) of reference \cite{DFF}. The solution is given by the relation:

\be
\label{eq:centodieci}
q^{2}(t)=2t^{2}H-4t D_{0}+2K_{0}.
\ee

\noindent As $(H,D_{0},K_{0})$ are constants of motion, once their values are assigned we
stick them in eq. (\ref{eq:centodieci}), and we get a relation between ``$q$" (on the LHS of 
(110) ) and ``$t$" on the RHS. This is the solution of the equation
of motion with ``initial conditions" given by the values we assign
to the constants of motion $(H,D_{0},K_{0})$. The reader may object
that we should give only 
two constant values (corresponding to the initial conditions $(q(0),{\dot q}(0))$) and not three.  
Actually the three values assigned to $(H,D_{0},K_{0})$ are not arbitrary because, as it was
proven in eq.(2-36) of ref.\scite{DFF}, these three quantities are linked
by a constraint:

\be
\label{eq:centoundici}
\left(HK_{0}-D_{0}^{2}\right)={g\over 4}
\ee

\noindent where ``$g$" is the coupling which entered the original Hamiltonian (see eq.(1)
of the present paper). Having one constraint among the three constants of motion brings them
down to two.

\noindent The proof of the relation (\ref{eq:centodieci}) above is quite simple. On the RHS, as the
$(H,D_{0},K_{0})$ are constants of motion, we can replace them with 
their time dependent expression $(H,D,K)$ (see eq.(89)-(91)), which are
explicitly:

\bea
\label{eq:centododici}
H & = & {1\over 2}\left({\dot q}^{2}(t)+{g\over q^{2}(t)}\right)\\
\label{eq:centotredici}
D & = & tH-{1\over 2}q(t){\dot q}(t)\\
\label{eq:centoquattordici}
K & = & t^{2}H-t~q(t){\dot q}(t)+{1\over 2}q^{2}
\eea

\noindent Inserting these expressions on the RHS of eq.(\ref{eq:centodieci}) we get immediately
the LHS. From the eqs.(112)-(114) it is also easy to see the relations
between the initial conditions $(q(0),{\dot q}(0))$ and the constants
$(H,D_{0},K_{0}$); in fact:

\bea
\label{eq:centoquindici}
H & = & H(t=0)={1\over 2}\left({\dot q}^{2}(0)+{g\over q^{2}(0)}\right)\\
\label{eq:centosedici}
D & = & D_{0}=-{1\over 2}\left(q(0){\dot q}(0)\right)\\
\label{eq:centodiciassette}
K & = & K_{0}={1\over 2}q^{2}(0).
\eea

\noindent From the relations above we see that, inverting them,
we can express ($q(0)$,${\dot q}(0)$) in term of $(H,D_{0},K_{0})$.
The constraint (111) is already taken care by the form the
$(H,D_{0},K_{0})$ have in terms of $(q(0)$,${\dot q}(0))$.

What we want to do in this section is to see if a relation
analogous to (110) exists also for our supersymmetric extension
or in general if the supersymmetric system can be solved exactly.
The answer is {\it yes} and it is based on a very simple trick.

Let us first remember eq. (78) which told us how $\HT$ and $H$ are related:

\be
\label{eq:centodiciotto}
i\int H(\Phi)~d\theta d{\bar\theta}=\HT
\ee

\noindent The same relation holds for $\DT_{0}$ and $\KT_{0}$ with respect to $D_{0}$
and $K_{0}$ as it is clear from the explanation given in the paragraph above eqs.(45)(46):

\bea
\label{eq:centodiciannove}
i\int D_{0}(\Phi)~d\theta d{\bar\theta}& = & \DT_{0}\\
\label{eq:centoventi}
i\int K_{0}(\Phi)~d\theta d{\bar\theta}& = & \KT_{0}.
\eea

\noindent Of course the same kind of relations holds for the explicitly time-dependent
quantities of eqs.(90)-(93):

\bea
\label{eq:centoventuno}
i\int D(\Phi)~d\theta d{\bar\theta}& = & \DT\\
\label{eq:centoventidue}
i\int K(\Phi)~d\theta d{\bar\theta}& = & \KT.
\eea

\noindent Let us now build the following quantity:

\be
\label{eq:centoventitre}
2t^2H(\Phi)-4tD(\Phi)+2K(\Phi)
\ee

\noindent This is functionally the RHS of eq.(110) with
the superfield $\Phi^{a}$ replacing the normal phase-space variable
$\phi^{a}$. It is then clear that the following relation holds:

\be
\label{eq:centoventiquattro}
\left(\Phi^{q}\right)^{2}=2t^2H(\Phi)-4tD(\Phi)+2K(\Phi)
\ee

\noindent The reason it holds is because, in the proof of the analogous one in $q$-space
(eq.(110)), the only thing we used was the functional form of the $(H,D,K)$
that was given by eq.(112)-(114). So that relation holds irrespective
of the arguments, $\phi$ or $\Phi$, which enter our functions provided that
the functional form of them remains the same. 

\noindent Let us first remember the form of $\Phi^{q}$ which appears\footnote{The
index $(.)^{q}$ is not a substitute for the index ``$a$" but it
indicates, as we said many times before, the first half of the indices ``$a$". Let us remember
in fact that the first half of the $\phi^{a}$ are just the configurational variables
$q^{i}$ which in our case of a 1-dim. system is just one variable.}
on the LHS of eq.(124):

\be
\label{eq:centoventicinque}
\Phi^{q}(t,\theta,\bar\theta)=q(t)+\theta~c^{q}(t)+{\bar\theta}\omega^{qp}{\bar c}_{p}(t)+i
{\bar\theta}\theta~\omega^{qp}\lambda_{p}(t).
\ee

\noindent Let us now expand in $\theta$ and ${\bar\theta}$ the LHS and RHS of
eq.(124) and compare the terms with the same power of $\theta$ and ${\bar\theta}$.

The RHS is

\be
\label{eq:centoventisei}
\left(\Phi^{q}\right)^{2}=q^{2}(t)+\theta[2q(t)c^{\scriptscriptstyle q}(t)]+{\bar\theta}
[2q(t){\bar c}_{\scriptscriptstyle p}(t)]+
{\bar\theta}\theta[2iq(t)\lambda_{\scriptscriptstyle p}(t)+2
c^{\scriptscriptstyle q}(t){\bar c}_{\scriptscriptstyle p}(t)].
\ee
\vskip .5cm

\noindent The LHS is instead:
\newpage 

\bea
\label{eq:centoventisette}
2t^2H(\Phi)-4tD(\Phi)+2K(\Phi) & = & 2t^2H(\phi)-4tD(\phi)+2K(\phi)+\nonumber\\
& + & \theta\left[ 2t^2 N^t_{\scriptscriptstyle H}-4t\ND^{t}+2\NK^{t}\right]+\nonumber\\
& - & {\bar\theta}\left[2t^2{\overline{N}^t_{\scriptscriptstyle H}}-4t{\overline\ND}^{t}+
2{\overline\NK}^{t}\right]+\nonumber\\
& + & i\theta{\bar\theta}\left[2t^2\HT-4t\DT+2\KT\right]
\eea

\noindent where the $\NH,{\overline\NH}$,$\ND^{t},\NK^{t}$ are defined in
{\bf TABLE 3} and in eqs.(95)(96),\break while the $\overline{N}^{t}_D$,
$\overline{N}^t_K$ are the time-dependent version\footnote{By ``time-dependent
version" we mean that they are related to the time independent ones
in the same manner as the $N^{t}$-functions were via eqs. (95)(98).} of the
operators defined\break in eq.(A.10). It is a simple exercise to
show that all these functions\break $(H,D,K,N_{(\ldots)},{\overline
N}_{(\ldots)},\HT,\KT,\DT)$ are conserved and constants of motion
in the enlarged space \quattrova.

If we now compare the RHS of eq.(126) and eq.(127) and equate terms with the
same power of $\theta$ and ${\bar\theta}$, we get (by writing the $N$ and
${\overline N}$ explicitly):

\bea
\label{eq:centoventotto}
q^{2}(t) & = & 2t^2H(\phi)-4tD(\phi)+2K(\phi);\\
2q(t)c^{q}(t) & = & \left[2t^2{\partial H\over\partial\phi^{a}}-4t{\partial
D\over\partial\phi^{a}}+2{\partial K\over\partial\phi^{a}}\right]c^{a};\\
2q(t){\bar c}_{p}(t) & = & \left[2t^2{\partial H\over\partial\phi^{a}}-4t{\partial
D\over\partial\phi^{a}}+2{\partial K\over\partial\phi^{a}}\right]\omega^{ab}
{\bar c}_{b};\\
i2q(t){\lambda_{\scriptstyle p}}(t)+2 c^{q}(t){\bar
c}_{p}(t)& = & -i\left[2t^2\HT-4t\DT+2\KT\right].
\eea

\noindent We notice immediately that eq.(128) is the same as the one of the original
paper\scite{DFF} and solves the motion for ``$q$". Given this solution
we plug it in eq.(129) and, as on the RHS we have the $N$-functions which
are constants, once these constants are assigned we get the motion of
$c^{\scriptscriptstyle q}$. Next
we assign three constant values to the ${\overline N}$-functions
which appear on the RHS of eq.(130), then we plug in the solution for $q$ given by 
eq.(128) and so we get the trajectory for ${\bar c}_{p}$. Finally we do the
same in eq.(131) and get the trajectory of $\lambda_{\scriptscriptstyle p}$.

The solution for the momentum-quantities 
$(p,c^{\scriptscriptstyle p},{\bar c}_{\scriptscriptstyle
q},\lambda_{\scriptscriptstyle q})$ can be obtained via their
definition in terms of the previous variables.

The reader may be puzzled by the fact that in the space \quattrova we have 8
variables but we have to give 12 constants of
motion: $(H,D_{0},K_{0},N_{(\ldots)},{\overline N}_{(\ldots)},\HT,\KT,\DT)$
to get the solutions from equations (128)-(131). The point is that, like in the case
of the standard conformal mechanics\scite{DFF}, we have constraints
among the constants of motion. We have already one constraint and it is
given by eq.(111). The others can be obtained in the following manner:
let us apply the operator $c^{\scriptscriptstyle a}
\partial_{\scriptscriptstyle a}$ on both sides of eq.(111) and what we get is the following relation:

\be
\label{eq:centotrentadue}
\NH K_{0}+\NK H-2\ND D_{0}=0
\ee 

\noindent which is a constraint for the $N$-functions. Let us now do the same applying
on both side of eq.(111) the operator ${\bar c}_{a}\omega^{ab}
\partial_{\scriptscriptstyle b}$. What we get is:

\be
\label{eq:centotrentatre}
{\overline \NH}K_{0}+{\overline \NK}H-2{\overline \ND}D_{0}=0
\ee

\noindent which is a constraint among the ${\overline N}$-functions. Finally let us
apply the $\Qb$ on equation (133) and we will get:

\be
\label{eq:centotrentaquattro}
i\HT K_{0}+i\KT H-21{\DT}D_{0}-{\overline \NH}
\NK-{\overline \NK}\NH+2{\overline \ND}\ND=0
\ee

\noindent which is a constraints among the $\HT,\DT,\KT$.

So we have 4 constraints (134)(133)(132)(111) which bring down
the constants of motion to be specified in \quattrova  from 12 to 8
as we expected.


\section{Conclusion}

In this paper we have provided a new supersymmetric extension of conformal
mechanics. We have realized that the model is deeply geometrical in the
sense that the Grassmannian variables and the 
supersymmetric Hamiltonian and various other charges
are all well-known objects in differential geometry.
In this case it is the differential geometry of the manifold and of the flows
associated to the original conformal mechanical model.
We feel it was important to unveil the geometry because the recently
discovered connection between conformal mechanics and black-holes
or in general the {\it Ads/CFT} connection must have deeply geometrical
origin.

What remains to be done, from a purely formal point of view, is to extend to
our model the recent analysis carried out in ref.\scite{KUM} where it was
found that the original conformal mechanics model has a Virasoro and
$w_{\scriptscriptstyle\infty}$ algebra. We hope to come back to these
issues using the tools developed in ref.\scite{MART}.

Last but not least, in performing this analysis we have discovered 
also a new {\it universal} symmetry present also for non-conformal model 
and which, in our opinion is at the heart of the interplay 
classical-quantum mechanics. 

\newpage

\begin{center}
{\LARGE\bf Appendix}
\end{center}

\appendix
\makeatletter
\@addtoreset{equation}{section}
\makeatother
\renewcommand{\theequation}{\thesection.\arabic{equation}}

\section{Details of the derivation of eqs.(84)-(87)}

In this appendix we are going to show the detailed calculations leading to
eqs.(\ref{eq:ottantaquattro})-(\ref{eq:ottantasette}).
The reader may have notice the  similarity between the charge
$\QH$ (\ref{eq:quaranta}) and the $\QD$,$\QK$ of eqs.(\ref{eq:quarantasette})
({\ref{eq:quarantotto}). We say ``similarity" because all of them are made of
two pieces, the first is the $\Qb$ for all of them. It is easy to show
that also the second pieces can be put in a similar form. Like for $\QH$ the second
piece was (see {\bf TABLE 3}) of the form $\NH=c^{a}\partial_{a}H$, so it is
easy to show that both $\QD$, and $\QK$ can be put in the form:

\setcounter{equation}{0}
\be
\QD=\Qb-2\gamma~\ND;~~~~~~~\QK=\Qb-\alpha~N_{k}
\ee 

\noindent
where $\ND$ and $\NK$ are respectively:

\setcounter{equation}{1}
\be
\ND=c^{a}\partial_{a}D_0;~~~~~~~
\NK=c^{a}\partial_{a}K_0
\ee

\noindent
with the $D_0$ and $K_0$ given\footnote{Actually we should take the classical
version of (17)(18) as we are doing classical mechanics.} by eqs. (\ref{eq:diciassette})(\ref{eq:diciotto}).
\noindent
So all three $(\NH,\ND, \NK)$ operators could be put in the general form:

\setcounter{equation}{2}
\be
N_{\scriptscriptstyle X}=c^{a}\partial_{a}X
\ee

\noindent
where $X$ is either $H,D_0$ or $K_0$. In the case of $D_0$ and $K_0$ the $X$ is quadratic
in the variables~$\phi^{a}$:

\setcounter{equation}{3}
\be
X={1\over 2}X_{ab}\phi^{a}\phi^{b}
\ee

\noindent
where $X_{ab}$ is a constant $2\times 2$ matrix.

In order to find the ${\widehat N}_{\scriptscriptstyle X}$ (that is the superspace
version of ${ N}_{\scriptscriptstyle X}$) we should use  eqs.~(79),(80) where 
$\Omega$ is now our operator $N_{\scriptscriptstyle X}$. From the expression of 
~$N_{\scriptscriptstyle X}$ we get for ~$\delta\Phi^{a}(t,\theta,{\bar\theta})$ of eq.(80):

\setcounter{equation}{4}
\be
\delta\Phi^{a}(t,\theta,{\bar\theta})={\bar\theta}\omega^{ab}({\bar\varepsilon}
\partial_{b}X)+i{\bar\theta}\theta\omega^{ab}(i\bar\varepsilon c^{d}\partial_{d}\partial_{b}X)
\ee

\noindent
where ${\bar\varepsilon}$ is the anticommuting parameter associated to the transformation.
\noindent
Given the form of $X$ (see eq.(A.4) above), we get for (A.5):

\setcounter{equation}{5}
\be
\delta\Phi^{a}(t,\theta,{\bar\theta})={\bar\theta}{\bar\varepsilon}\omega^{ab}X_{bd}[\phi^{d}+
\theta c^{d}].
\ee

\noindent
Note that, using superfields,  the above expression  can be written as:

\setcounter{equation}{6}
\be
\delta\Phi^{a}(t,\theta,{\bar\theta})=-{\bar\varepsilon}\omega^{ab}X_{bd}\bar{\theta}\Phi^{d}(t,\theta,
{\bar\theta}).
\ee

\noindent
So we obtain from eq.(79) that the superspace expression of $N_{\scriptscriptstyle X}$ is

\setcounter{equation}{7}
\be
({\widehat N}_{\scriptscriptstyle X})^a_d=\omega^{ab}X_{bd}{\bar\theta}.
\ee

\noindent The same kind of analysis we did here for the $\QD$ and $\QK$ can be done also
for the $\QBD$ and $\QBK$. They can be written as:

\setcounter{equation}{8}
\be
\QBD=\QBb+2\gamma~{\overline N}_{\scriptscriptstyle D};~~~~~ \QBK=\QBb+\alpha~{\overline
N}_{\scriptscriptstyle K};
\ee
\noindent
with 

\setcounter{equation}{9}
\be
{\overline N}_{D}={\bar c}_{a}\omega^{ab}\partial_{b}D;~~~~~~~~{\overline N}_{K}={\bar
c}_{a}\omega^{ab}\partial_{b}K;
\ee

\noindent
and the superspace representation of the ${\overline N}_{\scriptscriptstyle X}$
turns out to be:

\setcounter{equation}{10}
\be
({\widehat{\overline N}}_{\scriptscriptstyle X})^a_b=\omega^{ac}X_{cb}~\theta.
\ee

\noindent Remembering the form of the $D_0$ and $K_0$ functions in their classical version 
(see eq.(17)(18)) and comparing it with the general form of $X$ of eq.(A.4) above,
we get from eqs. (A.2)(A.3)  that the matrices $X_{ab}$  associated to $D$ and $K$ 
are\footnote{We will call this
form of $X_{ab}$ as $D_{ab}$, and the one associated to $K_0$ as $K_{ab}$, to stick to the
conventions of eqs.(84)-(87).} exactly those of eq.(88). So this is what we wanted
to prove.
\newpage
\section{Representation of {\bf\it H, D, K} in superspace}

In this appendix we will reproduce the calculations which provide
the superspace representations of $(H,D,K)$ contained in {\bf TABLE 11}.
We will start first with the operators at time $t=0$ which are listed in eqs.
(16)-(18). Using eqs.(79)(80) let us first do the variations
$\delta_{\scriptscriptstyle(H,D,K)}\Phi^{a}$. As $(H,D_{0},K_{0})$ 
contain only $(\phi^{a})$ their action will affect
only the $\lambda_{a}$ field contained in the superfield $\Phi$:
\setcounter{equation}{0}
\bea
\delta_{\scriptscriptstyle H}\lambda_{q}& = & \varepsilon
[H,\lambda_{q}]=-i\varepsilon{g\over q^{3}}\\
\delta_{\scriptscriptstyle H}\lambda_{p}& = & \varepsilon
[H,\lambda_{p}]=i\varepsilon p\\
\delta_{\scriptscriptstyle D_{0}}\lambda_{q}& = & \varepsilon
[D_{0},\lambda_{q}]=-{i\over 2}\varepsilon p\\
\delta_{\scriptscriptstyle D_{0}}\lambda_{p}& = & \varepsilon
[D_{0},\lambda_{p}]=-{i\over 2}\varepsilon q\\
\delta_{\scriptscriptstyle K_{0}}\lambda_{q}& = & \varepsilon
[K_{0},\lambda_{q}]=i\varepsilon q\\
\delta_{\scriptscriptstyle K_{0}}\lambda_{p}& = & \varepsilon
[K_{0},\lambda_{p}]=0.
\eea

\noindent Considering that the two superfields are:

\setcounter{equation}{6}
\bea
\Phi^{q}& = & q+\theta~c^{q}+{\bar\theta}{\bar c}_{p}+i{\bar\theta}\theta\lambda_{p};\\
\Phi^{p}& = & p+\theta~c^{p}-{\bar\theta}{\bar c}_{q}-
i{\bar\theta}\theta\lambda_{q};
\eea

\noindent
it is very easy to see that the ${\widehat\Omega}$ operators on the
RHS of eqs.(79) can only be the following:

\setcounter{equation}{8}
\bea
{\hat H} & = & {\bar\theta}\theta{\partial\over\partial t};\\
{\hat D}_{0} & = & -\frac{1}{2}{\bar\theta}{\theta}\sigma_{3};\\
{\hat K}_{0} & = & -{\bar\theta}\theta \sigma_{-}.
\eea
Next we should pass to the representation of the time-dependent
operators which are related to the time-independent ones by
eqs.(89)-(91). Also for the  superspace representation 
there will be the same relations between the two set of operators,
that means:

\setcounter{equation}{11}
\bea
{\hat H} &= & {\hat H}_{0};\\
{\hat D} & = & t {\hat H}+{\hat D}_{0};\\
{\hat K} & = & t^{2}{\hat H}+2t{\hat D}_{0}+{\hat K}_{0}.
\eea
\noindent
Using the above relations and the expressions obtained
in eqs.(B.8)-(B.10), it is easy to reproduce the last three
operators contained in {\bf TABLE 11}.

\section*{Acknowledgments}
We wish to thank  S.Minwalla and M.Reuter for useful correspondence.
The work contained here has been supported by grants from MURST (Italy) and
NATO. 
\newpage

\end{document}